\documentclass[iop,twocolumn]{aastex631}
\usepackage{mathrsfs}
\usepackage{amsmath}
\usepackage{float}
\usepackage{amstext}
\usepackage{gensymb}

\hypersetup{
   colorlinks,
   linkcolor={blue!88!black!80},
   citecolor={blue!88!black!80},
   urlcolor={blue!88!black!80}}

\newcommand{\ha}{H$\alpha$}
\newcommand{\hb}{H$\beta$}

\shorttitle{}

\turnoffeditone

\begin{document}

\title{The Intermediate-Mass Black Hole Reverberation Mapping Project: \\ Scientific Overview and Sample Characteristics}


\author[0000-0001-8416-7059]{Hengxiao Guo} 
\affiliation{Shanghai Astronomical Observatory, Chinese Academy of Sciences, 80 Nandan Road, Shanghai 200030, P. R. China}

\author[0000-0002-2581-8154]{Jiancheng Wu} 
\affiliation{Institute for Astronomy, School of Physics, Zhejiang University, 866 Yuhangtang Road, Hangzhou 310058, P. R. China}

\author[0000-0002-4521-6281]{Wenwen Zuo}
\affiliation{Shanghai Astronomical Observatory, Chinese Academy of Sciences, 80 Nandan Road, Shanghai 200030, P. R. China}

\author[0009-0009-0809-042X]{Ruining Tian}
\affiliation{Institute of Science and Technology for Deep Space Exploration, Suzhou Campus, Nanjing University, Suzhou 215163, P. R. China}

\author[0000-0002-1564-0436]{Xuechen Zheng}
\affiliation{Shanghai Astronomical Observatory, Chinese Academy of Sciences, 80 Nandan Road, Shanghai 200030, P. R. China} 

\author[0000-0001-9062-8309]{Meicun Hou}
\affiliation{Institute of Science and Technology for Deep Space Exploration, Suzhou Campus, Nanjing University, Suzhou 215163, P. R. China}

\author[0000-0003-1523-9164]{Paulina Lira}
\affiliation{Departamento de Astronom \`{\i}a, Universidad de Chile, Casilla 36D, Santiago, Chile}

\author[0000-0002-8186-4753]{Philip G. Edwards} 
\affiliation{CSIRO Space and Astronomy, PO Box 76, Epping, NSW, 1710, Australia}

\author[0000-0002-1912-0024]{Vivian U}
\affiliation{IPAC, Caltech, 1200 E. California Blvd., Pasadena, CA 91125, USA}
\affiliation{Department of Physics and Astronomy, 4129 Frederick Reines Hall, University of California, Irvine, CA 92697, USA}

\author[0000-0002-2052-6400]{Shu Wang}
\affiliation{School of Physics \& Astronomy, University of Southampton, Southampton, SO17 1BJ, United Kingdom} 

\author[0000-0003-4440-259X]{Mar Mezcua}
\affiliation{Institute of Space Sciences (ICE, CSIC), Campus UAB, Carrer de Magrans, 08193 Barcelona, Spain}
\affiliation{Institut d'Estudis Espacials de Catalunya (IEEC), Edifici RDIT, Campus UPC, 08860 Castelldefels (Barcelona), Spain} 

\author[0000-0001-6947-5846]{Luis C. Ho}
\affiliation{Department of Astronomy, School of Physics, Peking University, Beijing 100871, P. R. China} 
\affiliation{Kavli Institute for Astronomy and Astrophysics, Peking University, Beijing, 100871, P. R. China}

\author[0000-0002-4455-6946]{Minfeng Gu}
\affiliation{Shanghai Astronomical Observatory, Chinese Academy of Sciences, 80 Nandan Road, Shanghai 200030, P. R. China}

\author[0000-0003-4341-0029]{Tao An}
\affiliation{Department of Astronomy, University of Science and Technology of China, 96 Jinzhai Road, Hefei, Anhui 230026, China}

\author[0000-0002-5248-2422]{Samuzal Barua}
\affiliation{Shanghai Astronomical Observatory, Chinese Academy of Sciences, 80 Nandan Road, Shanghai 200030, P. R. China}

\author[0000-0001-9947-6911]{Colin J. Burke}
\affiliation{Department of Physics, University of North Texas, Denton, TX 76203, USA}

\author[0000-0002-4223-2198]{Zhen-yi Cai}
\affiliation{Department of Astronomy, University of Science and Technology of China, 96 Jinzhai Road, Hefei, Anhui 230026, China}

\author[0000-0001-8917-2148]{Xuheng Ding}
\affiliation{School of Physics and Technology, Wuhan University, Wuhan 430072, P. R. China}

\author[0000-0002-1530-2680]{Haicheng Feng}
\affiliation{Yunnan Observatories, Chinese Academy of Sciences, 396 Yangfangwang, Guandu District, Kunming 650216, Yunnan, P. R. China}

\author[0000-0002-9331-4388]{Alok C. Gupta}
\affiliation{Aryabhatta Research Institute of Observational Sciences (ARIES), Manora Peak, Nainital 263001, India} 

\author[0000-0003-3823-3419]{ShaSha Li}
\affiliation{Yunnan Observatories, Chinese Academy of Sciences, 396 Yangfangwang, Guandu District, Kunming 650216, Yunnan, P. R. China}

\author[0009-0007-1193-609X]{Wanling Liu}
\affiliation{Shanghai Astronomical Observatory, Chinese Academy of Sciences, 80 Nandan Road, Shanghai 200030, P. R. China}
\affiliation{University of Chinese Academy of Sciences, 19A Yuquan Road, 100049, Beijing, P. R. China}

\author[0000-0002-1820-7865]{Wen-juan Liu}
\affiliation{Yunnan Observatories, Chinese Academy of Sciences, 396 Yangfangwang, Guandu District, Kunming 650216, Yunnan, P. R. China}
\affiliation{Key Laboratory for the Structure and Evolution of Celestial Objects, Chinese Academy of Sciences, Kunming 650216, Yunnan, P. R. China}
\affiliation{Center for Astronomical Mega-Science, Chinese Academy of Sciences, 20A Datun Road, Chaoyang District, Beijing 100012, P. R. China}

\author[0000-0002-7692-7967]{Ru-sen Lu}
\affiliation{Shanghai Astronomical Observatory, Chinese Academy of Sciences, Shanghai 200030,  P. R. China}
\affiliation{Key Laboratory of Radio Astronomy and Technology, Chinese Academy of Sciences, A20 Datun Road, Chaoyang District, Beijing, 100101, P. R. China}
\affiliation{Max-Planck-Institut f\"ur Radioastronomie, Auf dem Hf\"urgel 69, D-53121 Bonn, Germany}

\author[0000-0002-1134-4015]{Dragana Ili\'{c}}
\affiliation{University of Belgrade - Faculty of Mathematics, Department of Astronomy, Studentski trg 16, Belgrade, Serbia}
\affiliation{Hamburger Sternwarte, Universitat Hamburg, Gojenbergsweg 112, D-21029 Hamburg, Germany}

\author[0000-0001-5139-1978]{Andjelka B. Kova\v{c}evi\'{c}}
\affiliation{University of Belgrade - Faculty of Mathematics, Department of Astronomy, Studentski trg 16, Belgrade, Serbia}

\author[0009-0002-7625-2653]{Yu Pan}
\affiliation{South-Western Institute for Astronomy Research, Key Laboratory of Survey Science of Yunnan Province, Yunnan University, Kunming, Yunnan 650504, P. R. China}

\author[0000-0003-2398-7664]{Luka \v{C}. Popovi\'{c}}
\affiliation{University of Belgrade - Faculty of Mathematics, Department of Astronomy, Studentski trg 16, Belgrade, Serbia}
\affiliation{Astronomical Observatory, Volgina 7, 11060 Belgrade, Serbia}

\author[0000-0002-3742-6609]{Wenke Ren}
\affiliation{Shanghai Astronomical Observatory, Chinese Academy of Sciences, 80 Nandan Road, Shanghai 200030, P. R. China}

\author[0000-0003-0820-4692]{Paula S\'{a}nchez-S\'{a}ez}
\affiliation{European Southern Observatory, Karl-Schwarzschild-Str. 2, 85748, Garching, Germany}

\author[0000-0001-8416-7059]{Jamie Stevens}
\affiliation{CSIRO Astronomy and Space Science, PO Box 76, Epping, NSW, 1710, Australia}

\author[0000-0001-8416-7059]{Jingbo Sun} 
\affiliation{Shanghai Astronomical Observatory, Chinese Academy of Sciences, 80 Nandan Road, Shanghai 200030, P. R. China}
\affiliation{University of Chinese Academy of Sciences, 19A Yuquan Road, 100049, Beijing, P. R. China} 

\author[0000-0002-0771-2153]{Mouyuan Sun}
\affiliation{Department of Astronomy, Xiamen University, Xiamen, Fujian 361005, P. R. China}

\author[0009-0002-5955-4932]{Chizhuo Wang}
\affiliation{Department of Astronomy, School of Physics, Peking University, Beijing 100871, P. R. China} 
\affiliation{Kavli Institute for Astronomy and Astrophysics, Peking University, Beijing, 100871, P. R. China}

\author[0000-0002-4419-6434]{Junxian Wang}
\affiliation{Department of Astronomy, University of Science and Technology of China, 96 Jinzhai Road, Hefei, Anhui 230026, P. R. China} 
\affiliation{College of Physics, Guizhou University, Guiyang, Guizhou, 550025, P. R. China}

\author[0000-0001-5012-2362]{Rongfeng Shen}
\affiliation{School of Physics and Astronomy, Sun Yat-Sen University, Zhuhai, 519082, P. R. China}
\affiliation{CSST Science Center for the Guangdong-Hongkong-Macau Greater Bay Area, Sun Yat-Sen University, Zhuhai, 519082, P. R. China}

\author[0000-0002-7350-6913]{Xuebing Wu}
\affiliation{Department of Astronomy, School of Physics, Peking University, Beijing 100871, P. R. China} 
\affiliation{Kavli Institute for Astronomy and Astrophysics, Peking University, Beijing, 100871, P. R. China}

\author[0000-0002-8614-6275]{Yong Shi}
\affiliation{Department of Astronomy, Westlake University, Hangzhou, Zhejiang 310030, P. R. China} 

\author[0000-0003-0644-9282]{Zhefu Yu}
\affiliation{Kavli Institute for Particle Astrophysics and Cosmology (KIPAC), Stanford University, Stanford CA 94305, USA}

\author[0000-0002-9634-2923]{Zhenya Zheng}
\affiliation{Shanghai Astronomical Observatory, Chinese Academy of Sciences, 80 Nandan Road, Shanghai 200030, P. R. China} 

\author[0000-0002-8005-0870]{Ling Zhu}
\affiliation{Shanghai Astronomical Observatory, Chinese Academy of Sciences, 80 Nandan Road, Shanghai 200030, P. R. China}

\begin{abstract}
Recent discoveries with the {\it James Webb Space Telescope} of massive black holes at high redshift have highlighted fundamental questions about black hole seed formation and the coevolution of black holes with their host galaxies. Because the initial seed population cannot yet be observed directly, nearby intermediate-mass black holes provide a complementary fossil record of black hole formation and early growth. Motivated by this opportunity, we present the Intermediate-Mass Black Hole Reverberation Mapping (IMBH-RM) project and construct a homogeneous Sloan Digital Sky Survey sample of active broad-line IMBHs by uniformly reanalyzing literature candidates with consistent spectral decomposition and black hole mass estimation. Our sample contains 192 reliable IMBH candidates at $z\lesssim0.3$ with $\log(M_{\rm BH}/M_\odot)<6$, including four particularly compelling sources with $\log(M_{\rm BH}/M_\odot)<5$. The primary goal of IMBH-RM is to obtain reliable black hole masses from direct measurements and characteristic sizes of the broad-line region and accretion disk for a carefully selected subsample. These measurements will provide robust low-mass anchors for calibrating single-epoch black hole mass estimates and extending black hole--galaxy scaling relations into the IMBH regime. By building a statistically meaningful reverberation-mapped sample spanning $10^4$--$10^6\,M_\odot$, we aim to constrain the local IMBH mass distribution and place observational constraints on competing black hole seed formation scenarios. The future Multi-Channel Imager aboard the Chinese Space-station Survey Telescope provides a particularly promising platform for achieving these goals.

\end{abstract}


\section{Introduction}\label{sec:intro}
\subsection{Black Hole Seeds and the Assembly of Supermassive Black Holes}

\begin{figure*}[!t]
    \centering
    \includegraphics[width=1\linewidth]{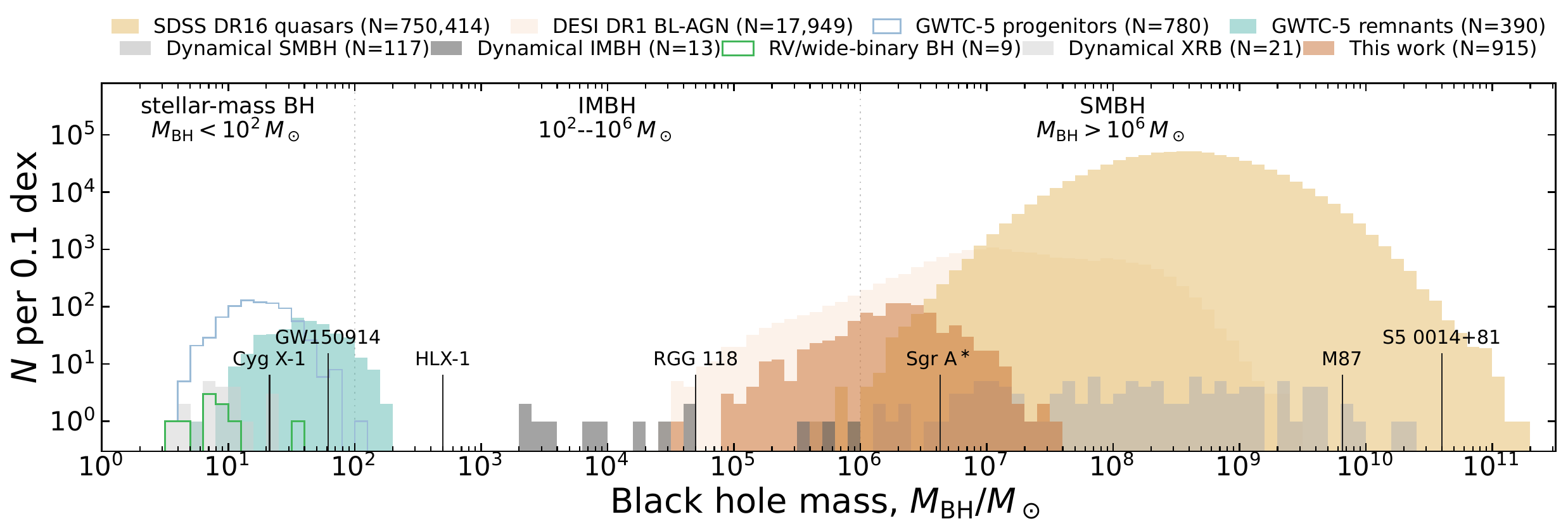}
    \caption{Observed black-hole mass distribution. The stellar-mass regime includes dynamically measured black holes in X-ray binaries from \citet{Corral-Santana16}, GWTC-5 binary-component and merger-remnant masses \citep{GWTC5}, and secure RV/astrometric wide-binary black holes compiled by \citet{WangS24}. At the supermassive end, we show SDSS DR16 quasar single-epoch virial masses \citep{Wu22} and deduplicated dynamical measurements from \citet{Graham23} and \citet{Saglia16} as representative samples. The orange histogram represents the low-mass black holes identified in this work, substantially filling the sparsely populated IMBH regime. Previously reported dynamical IMBH measurements are adopted from the compilation of \citet{Burke26} and the references therein. The DESI DR1 broad-line AGN sample from \citet{Pucha26}, shown in light orange, contains both IMBH and SMBH samples. We also mark several well-known black holes for reference: the first GW-detected binary merger, GW150914 \citep{Abbott16}; the low-mass dwarf AGN, RGG~118 \citep{Baldassare15}; the hyperluminous IMBH candidate, HLX-1 \citep{Farrell09}; the Galactic-center SMBH, Sgr~A* \citep{Ghez08}; the first EHT-imaged SMBH, M87 \citep{EHT19}; and the ultramassive blazar, S5~0014+81 \citep{Ghisellini09}. }
    \label{fig:BH_Census}
\end{figure*}

Black holes span a remarkable range of masses, from stellar-mass black holes with $\lesssim 10^2\,M_\odot$ to supermassive black holes (SMBHs) with $\sim10^{6-10}\,M_\odot$. Stellar-mass black holes are primarily identified through X-ray binary observations and gravitational-wave detections. The former has confirmed about 20 sources in the Milky Way (e.g., the BlackCAT catalog; \citealt{Corral-Santana16}), while the LIGO--Virgo--KAGRA network has detected hundreds of compact-binary mergers, predominantly involving black holes. The latest GWTC-5.0 catalog brings the cumulative number of cataloged compact-binary coalescences to 390 through the fourth observing run (O4b; \citealt{GWTC5}). At the high-mass end, modern optical spectroscopic surveys, such as the Sloan Digital Sky Survey \citep[SDSS,][]{Ahumada20} and the Dark Energy Spectroscopic Instrument \citep[DESI,][]{DESI16}, have obtained spectra for millions of galaxies and quasars, enabling statistical studies of SMBH demographics and evolution. Yet, the intermediate-mass black hole (IMBH, $10^2-10^6\,M_\odot$\footnote{We adopt the historical definition of IMBHs \citep{Greene04,Mezcua17}.}) class, which bridges the stellar and supermassive scales, remains largely unexplored (see Figure \ref{fig:BH_Census}), leaving a critical gap in our understanding of black hole formation and growth.

As new observations from the \textit{James Webb Space Telescope} (JWST) continue to uncover first-generation galaxies and quasars at ever earlier cosmic epochs \citep[e.g., $z > 6$,][]{Kokorev23, Bunker23, Naidu25}, understanding how SMBHs formed so rapidly has become an increasingly pressing question \citep{Maiolino24, Carniani24}. To explain the presence of billion-solar-mass black holes less than a billion years after the Big Bang, two leading theoretical pathways have been proposed: the light- and heavy-seed scenarios \citep[see reviews by][]{Woods19,Inayoshi20,Volonteri21}.  In the light-seed scenario, the first black holes originate from the remnants of massive Population III stars \citep[e.g.,][]{Madau01}, with initial masses of $10^2–10^3\,M_\odot$. Growing such seeds into SMBHs observed in the infant Universe generally requires sustained accretion near or above the Eddington limit. In contrast, the heavy-seed scenario \citep[e.g.,][]{Bromm03,Begelman06} posits the direct collapse of primordial gas clouds into $10^4–10^6\,M_\odot$ black holes, allowing the formation of early SMBHs through more moderate, sub-Eddington growth rates (see reviews by \citealt{Greene20}).

Probing black hole seeds in the early Universe remains beyond current observational capabilities. Fortunately, theoretical work suggests that observations of local black holes in dwarf galaxies \citep[i.e., stellar mass $M_\star < 10^{9.5}\,M_\odot$,][]{Reines13} may offer crucial clues to understanding seeding processes, as these black holes have experienced relatively fewer mergers and less accretion since their formation compared to their supermassive counterparts \citep{Mezcua17,Volonteri21}. The key challenge in assessing seeding scenarios lies in building a robust statistical sample of IMBHs, as neither identifying them in low-mass dwarfs nor determining their nature through accurate mass measurements is trivial.

Black hole formation theories suggest that the observed black hole mass distribution in the local Universe contains valuable information about their initial seeding mechanisms \citep{Greene20,Volonteri21}. If light seeds dominated in the early Universe, we would expect numerous black holes around $10^{2-3}\,M_\odot$ today, whereas if heavy seeds prevailed, the population would instead be fewer but more massive ($10^{4-5}\,M_\odot$). These differences are predicted to be particularly evident at the low-mass end of the observed scaling relations (e.g., $M_{\rm BH}$–$\sigma_\star$ and $M_{\rm BH}$–$M_\star$): a prevalence of heavy seeds would flatten these relations at lower masses, while light seeds would preserve the slope, extending it toward lower-mass regimes. Accurately measuring the black hole mass function in this range thus provides a promising avenue for distinguishing among competing seeding scenarios, albeit with challenges and uncertainties \citep{Regan24,Matthee24,Greene24,Bhwmick25}. Nevertheless, building a robust IMBH sample remains essential, as it will bridge the gap between stellar-mass and supermassive black holes and provide crucial insights into their demographics and growth histories.

A complementary approach is to identify individual high-redshift galaxies that are extreme outliers from the $M_{\rm BH}$–$M_{*}$ relation. Whereas statistical measurements of the relation constrain the average coevolution of black holes and their hosts, individual overmassive or undermassive systems may provide more direct snapshots of their relative growth at early cosmic times, although they may not be representative of the overall population. In particular, a strongly overmassive black hole in a still-assembling galaxy may be indicative of a massive seed that formed and became active before substantial stellar growth, while a genuinely undermassive black hole may favor a lighter seed whose subsequent growth lagged behind that of its host. Recent JWST observations have revealed several candidate overmassive black holes that have been interpreted as possible signatures of heavy seeds \citep{Kokorev23,Maiolino24,Larson23}, although their inferred offsets may also be affected by selection biases and uncertainties in black hole and stellar-mass estimates \citep{Ding23,Li25,Silverman25}. Such individual systems cannot by themselves uniquely identify the dominant black hole seeding channel, but extreme, well-characterized outliers can provide valuable constraints on seeding models and complement statistical studies of black hole–galaxy scaling relations.

\subsection{Reverberation Mapping}
The most direct way to establish the presence of an IMBH is to dynamically measure the central black hole mass using stellar or gas kinematics. However, because the sphere of influence of an IMBH is exceedingly compact, such measurements are generally feasible only within the Local Volume ($\lesssim 10$ Mpc), where stellar or gas kinematics can still be spatially resolved with current instruments \citep{denBrok15,Nguyen18}. Reverberation mapping (RM) offers a powerful alternative by probing the central regions through light echoes from the broad-line region (BLR) surrounding the black hole \citep{Blandford82,Peterson93,Peterson14}.

Over the past three decades, the RM technique has been successfully applied to measure black hole masses in over 200 SMBHs \citep[see the compilation in ][]{Wang24}. The pioneering international AGN Watch project conducted extensive monitoring campaigns of several AGNs, including NGC~5548, firmly establishing RM as a robust method for probing the size, structure, and kinematics of the BLR around SMBHs \citep{Clavel91, Peterson91}, and deriving the foundational BLR radius–luminosity ($R$–$L$) relationship \citep{Kaspi00,Bentz13}, which enables mass estimates for large AGN samples from single-epoch spectroscopy.

Subsequently, multiple groups significantly expanded the RM sample \citep{Bentz09,Bentz10,Grier12,Grier17b,Wang14,Barth11,Barth15,U22,Woo15,Woo24,Du18,Shen15,Shen24,Malik23,McDougall25}, exploring the universality of the $R$–$L$ relation across a wide range of AGN properties. Later studies revealed significant deviations from the \(R\)–\(L\) relation, particularly among rapidly accreting AGNs \citep[e.g.,][]{Du18,Du19,Wang24}. Such scatter may also arise from wavelength-dependent continuum delays and multicomponent BLR structures, which can bias the measured emission-line lags \citep{Feng24,LiSS26}. Another major advancement has been the development of BLR dynamical modeling, which interprets velocity-resolved RM data within a physical framework to infer the black hole mass and the geometry and kinematics of the BLR without relying on an externally calibrated virial factor. The pioneering studies of \citet{Pancoast11,Pancoast14} demonstrated the feasibility of this approach and paved the way for subsequent refinements \citep[e.g.,][]{Grier17a,Li18,Williams18,Villafana22,Wang26}. Complementarily, VLTI/GRAVITY spatially resolves velocity-dependent BLR photocenter shifts, constraining its size, orientation, and kinematics \citep{Gravity18,Gravity21}. Combined with RM, this spectroastrometry and reverberation mapping (SARM) approach enables joint measurements of black hole masses and geometric distances \citep{WangJM20,LiYR25}.

In addition to broad-line RM, continuum RM (CRM) offers a complementary approach to probing AGN accretion disks by measuring the wavelength-dependent continuum lags ($\tau$) predicted by the standard thin disk model \citep{Shakura73}, where $\tau \propto \lambda^{4/3}$. Early work detected such inter-band lags in nearby AGNs \citep[e.g.,][]{Collier98, Sergeev05}, motivating systematic campaigns. For example, the AGN STORM project \citep{Edelson15, Fausnaugh16} carried out the first intensive, multiwavelength CRM of NGC~5548, revealing larger-than-expected disk lags and suggesting that diffuse continuum emission from the BLR may contribute to this excess \citep[see also][]{Cackett18}. Similar findings have been reported in both single-object studies \citep[e.g.,][]{Edelson19,HS20} and larger samples \citep[e.g.,][]{Jiang17, Mudd18, Yu20,GuoWJ22}. Recently, \citet{Guo22} demonstrated that continuum lags and emitting region follow a tight radius–luminosity relation, supporting the scenario where BLR diffuse continuum dominates the observed lag signal. Building on this, \citet{Wang23} proposed using optical continuum lags as proxies for H$\beta$ lags to estimate black hole masses, finding that the 5100~\AA\ continuum-emitting region is typically a factor of $\sim$8 smaller than the H$\beta$ BLR \citep[see also][]{Mandal25}. 

Extending RM into the IMBH regime, the prototypical low-mass AGN NGC 4395 has served as a benchmark, demonstrating that both BLR and continuum RM can be successfully conducted on minute-to-hour timescales. Both BLR and continuum RM have been successfully carried out for NGC 4395 using spectroscopic and photometric monitoring across a broad wavelength range \citep{Peterson05,Cho20,Pandey24,Desroches06,Montano22,McHardy23,Sun25}. For example, previous campaigns measured an H$\alpha$ BLR lag of $83\pm14$ minutes \citep{Woo19} and optical $g$--$z$ continuum lags of approximately 10--20 minutes \citep{Montano22,Pan26}. These results provide a compelling proof of concept for RM across the IMBH mass range of $10^{4}$--$10^{6}\,M_\odot$. Motivated by this success, we designed a dedicated RM project to systematically expand lag and black hole mass measurements for IMBHs.

This work reviews broad-line IMBH candidates identified in the SDSS era and compiles published samples for a uniform reanalysis of their demographic and multiwavelength properties. The remainder of this paper is organized as follows. \S~\ref{sec:overview} introduces the motivation, goals, and observing strategy of the IMBH-RM project. \S~\ref{sec:sample} reviews previous IMBH samples, while \S~\ref{sec:results} presents our uniform reanalysis and characterization of IMBH candidates, including their multiwavelength properties and prospects for RM. Some initial RM results have been published in \citet{Sun25} and \citet{Pan26}. \S~\ref{sec:diss} discusses the main implications, and \S~\ref{sec:conclusion} summarizes our conclusions and future prospects. Throughout this paper, we adopt a flat $\Lambda$CDM cosmology with $H_0=70.0~{\rm km~s^{-1}~Mpc^{-1}}$ and $\Omega_{\rm m}=0.3$.

\section{Overview of IMBH-RM Project}\label{sec:overview}

\subsection{Motivations}
As outlined above, previous studies have established the technical feasibility of extending RM into the IMBH regime. Scientifically, recent JWST discoveries of massive black holes in the early Universe have heightened the urgency of understanding black hole formation and black hole--galaxy coevolution. Motivated by this technical foundation, these scientific imperatives, and our fundamental curiosity about the largely unexplored IMBH population bridging stellar-mass and supermassive black holes, we designed the IMBH-RM project to systematically investigate the $10^{4}$--$10^{6}\,M_\odot$ black hole mass regime.

\subsection{Scientific Goals}
The scientific goals of the IMBH-RM project are centered on three main pillars:
\begin{itemize}

\item Obtaining robust reverberation-based black hole masses for a carefully selected subsample of IMBH candidates, providing secure anchors at the low-mass end of the black hole population. These measurements will establish whether individual systems are overmassive or undermassive relative to their host galaxies and thereby provide constraints on black hole seeding scenarios.

\item Using these reverberation-based anchors to extend and calibrate the BLR radius--luminosity relation into the low-mass, low-luminosity regime, enabling reliable single-epoch mass estimates for much larger IMBH samples. The resulting black hole mass census will extend the $M_{\rm BH}$--$M_{*}$ and $M_{\rm BH}$--$\sigma_{*}$ relations to lower masses, providing statistical constraints on black hole seed formation and black hole--galaxy coevolution.

\item Measuring continuum reverberation lags across multiple wavelengths to map the structure of accretion disks in the IMBH regime. These observations will enable us to test the standard thin-disk model \citep{Shakura73}, X-ray reprocessing models \citep{Krolik91,Cackett07}, and other potential scenarios \citep{Gardner2017,Cai18,Sun20,Netzer2022}, while investigating the physical drivers of AGN variability and assessing continuum RM as an independent method for estimating black hole masses \citep{Wang23}.

\end{itemize}

\subsection{Sample and Observational Strategy}
Our parent sample is drawn from previously reported IMBH candidates in the literature, all identified based on single-epoch spectroscopic measurements showing broad Balmer emission lines and estimated black hole masses of $< 10^6$~M$_{\odot}$ (see \S~\ref{sec:sample}). From this parent sample, we will select approximately 10--20 of the most robust IMBH candidates, spanning a broad range in luminosity and black hole mass, for dedicated RM campaigns with ground- and space-based facilities. The selection will be guided by multi-wavelength evidence, including repeated spectroscopic confirmation of broad emission lines over a decade (Wu et al. in prep.), long-term variability across multiple bands (Liu et al., in prep.), and X-ray and radio properties.

For IMBHs with black hole masses below $10^{6}\,M_\odot$, broad-line lags (e.g., H$\beta$ and H$\alpha$) are generally expected to be shorter than $\sim$2 days, while optical continuum lags are typically below 1 day. Here, we use broad-line RM as an illustrative example to describe the observing strategy. The expected broad-line lags can be roughly estimated from empirical lag--luminosity relations \citep{Cho23,Dallabonta25}, enabling tailored monitoring campaigns. Based on the predicted reverberation timescales, targets are classified into three categories: short-, intermediate-, and long-lag systems.

Targets with long lags ($\gtrsim$ 1 day) require daily or sub-daily cadence sustained over 1–2 months, as demonstrated in campaigns of NGC~4051 \citep{Denney09} and UGC~06728 \citep{Bentz16}, which represent the lowest black hole masses in the SMBH regime. To ensure sufficient cadence, one or two ground-based telescopes at different sites with supplement visibility are often necessary to mitigate the impact of poor weather and other interruptions. Both spectroscopic and photometric monitoring are feasible for such long-lag targets, depending on source redshift, brightness and broad-line strength.

Targets with short lags ($\lesssim$ 1 night, 8 hr), such as NGC~4395, require continuous monitoring over an entire night with minute-level cadence using broadband and narrow-band photometry to resolve the rapid variability \citep{Woo19}. Conducting spectroscopic RM for such systems is substantially more demanding, as it requires high signal-to-noise spectra obtained at similarly high temporal cadence, often necessitating large-aperture facilities \citep{Cho20,Pandey24}.

Intermediate-lag targets (1 night to 1 day, 8--24 hr) are arguably the most challenging class for RM observations, as they often require relay monitoring across multiple longitudes or nearly continuous campaigns extending over several weeks. Space-based facilities such as the Chinese Space-station Survey Telescope \citep[CSST,][]{CSST} are particularly advantageous for this regime. As the CSST is in low Earth orbit, with a period of $\sim$90 minutes, there would be regular interruptions in the coverage for much (though perhaps not all) of the sky.
Alternatively, polar sites during the polar night offer a unique opportunity for continuous ground-based monitoring \citep{Yang23,Zhou26} owing to their prolonged darkness. Under these observing configurations, both spectroscopic and photometric RM measurements become practical for intermediate-lag systems.

\subsection{Difficulties and Opportunities}

Compared with SMBHs, IMBH-RM faces two main challenges. First, although the intrinsic variability of IMBHs may not be weak, these sources become fainter at larger distances, and their AGN variability is often strongly diluted by host-galaxy light, making significant variability difficult to detect within the available sensitivity and monitoring duration. Second, while the expected reverberation lags can be very short, the characteristic variability timescales may be much longer. For example, an active IMBH like POX 52 with a BLR lag of only $\sim0.5$ day (Zuo et al. in prep.) may have a characteristic variability timescale of a few days. Detecting significant variability therefore requires a monitoring baseline comparable to this characteristic timescale, whereas resolving the short lag simultaneously demands very high cadence. This combination makes IMBH-RM highly time-consuming and observationally expensive.

On the positive side, black holes in other nearby dwarf galaxies may become active through gas accretion, providing ideal new targets for IMBH-RM. Wide-field time-domain surveys, particularly the Vera C. Rubin Observatory Legacy Survey of Space and Time (LSST; \citealt{Ivezic19}), will greatly expand the pool of variable low-mass AGNs \citep[e.g.,][]{Bunker23}. The most promising sources can then be selected for dedicated high-cadence follow-up with CSST and, in the longer term, the Habitable Worlds Observatory (HWO; \citealt{Feinberg26}).

\begin{deluxetable}{lc}[ht]
\centering
\tabcolsep=0.4cm
\tablecaption{Number of IMBH candidates with $\log(M_{\rm BH}/M_\odot)\leq6.5$ reported in previous studies, listed in chronological order. Sources already included in earlier studies are excluded from each subsequent study. The final combined sample contains 1,447 unique sources.\label{tab:sample}}
\tablehead{
\colhead{Literature} & \colhead{Number}
}
\startdata
\cite{Kunth87} & 1 \\
\cite{Filippenko89} & 1 \\
\cite{Greene07} & 229 \\
\cite{Denney09} & 1 \\
\cite{Dong12} & 160 \\
\cite{Reines13} & 3 \\
\cite{Ho16} & 11 \\
\cite{Bentz16} & 1 \\
\cite{LiuH18} & 202 \\
\cite{Chilingarian18} & 293 \\
\cite{Mezcua20} & 1 \\
\cite{Salehirad22} & 29 \\
\cite{Shin22} & 14 \\
\cite{Lin24} & 47 \\
\cite{Mezcua24} & 7 \\
\cite{Bernal25} & 35 \\
\cite{LiuW25} & 412 \\
\hline
Total & 1,447 \\
\enddata
\end{deluxetable}

\section{Sample Characteristics} \label{sec:sample}
To measure reverberation-based black hole masses, we select only active IMBH candidates with broad emission lines. We primarily focus on targets with single-epoch black hole masses (virial estimates from line widths and luminosities from a single spectrum) of $< 10^{6}\,M_{\odot}$, but do not exclude objects with slightly higher masses, which serve as a useful comparison sample. This mass range ($10^{5-6}\,M_{\odot}$) is also expected to retain the strongest signatures of black hole seeding \citep{Bhowmick25}. The IMBH candidate sample is compiled from previous literature listed in Table \ref{tab:sample}, which provides a summary of the discoveries made during the SDSS era. We emphasize that our goal is to identify the most robust IMBH candidates, i.e., sources with reliable broad-line detections and black hole masses $< 10^{6}\,M_{\odot}$, rather than to construct a complete IMBH sample.

\subsection{Two prototypical IMBHs: NGC 4395 and POX 52} \label{sec:NGC4395}
NGC\,4395, first identified as an active nucleus in the late 1980s, is a nearby, bulgeless dwarf spiral galaxy hosting the least luminous known Seyfert\,1 \citep{Filippenko89,Filippenko1993}. It exhibits clear AGN variability from X-ray to optical bands \citep[e.g.,][]{Lira99,Iwasawa00,Burke20,Pan26}. RM was initiated by \citet{Peterson05,Desroches06}, who detected extremely short optical lags and estimated the black hole mass to lie in the intermediate-mass range. Subsequent continuum and broad-line RM campaigns \citep{Minezaki06,Woo19,Montano22,McHardy23,Sun25} consistently confirmed the short lags and low mass. Independent stellar dynamical modeling \citep{denBrok15,Brum19} further corroborated the RM results. Together, the multi-wavelength evidence, RM measurements, and dynamical modeling firmly establish NGC\,4395 as a prototypical IMBH AGN, with a black hole mass securely in the range of $\sim 10^4$--$10^5\,M_\odot$.

POX\,52, initially discovered in the early 1980s as a dwarf galaxy with Seyfert-like emission lines \citep{Kunth81,Kunth87}, was later confirmed to host a Seyfert\,1 nucleus with broad permitted lines and a low stellar velocity dispersion \citep{Barth04}. Multi-wavelength observations have revealed clear X-ray and radio variability \citep{Kawamuro24,Yuan24}, providing strong evidence for accretion activity. Recently, \citet{Sun25} measured an optical--MIR lag of $\sim40$ days. Converting this dust lag to an expected H$\beta$ lag and combining it with the broad-line width yields $\log(M_{\rm BH}/M_\odot)\sim5.5$, independently supporting the classification of POX~52 as an IMBH.

\subsection{Greene \& Ho Sample}
A robust IMBH sample is essential for studying their demographics, growth, and early coevolution with host galaxies. Using the SDSS database, \citet{Greene04,Greene07} systematically selected broad-line AGNs with virial black hole masses $\lesssim 2\times10^6\,M_\odot$, identifying 19 objects in their initial sample in DR1 and expanding to 174 robust IMBH candidates (and 55 less secure candidates) in DR4 (hereafter GH07). The candidates were identified based on the presence of broad H$\alpha$ emission and a virial mass estimate derived from the broad-line width and AGN luminosity. The sources extend to redshifts of $z\sim0.35$ and are hosted by late-type, low-luminosity galaxies. This sample provides a unique opportunity to study low-mass black holes as local analogs of primordial seeds and to 
inform studies of the black hole occupation fraction in dwarf galaxies. 

Multiwavelength follow-up observations have provided broad support for the IMBH interpretation of the GH07 sample \citep{Greene07}. Using improved optical spectroscopy, \citet{Barth2005} and \citet{Xiao11} obtained reliable stellar velocity dispersions for 71 objects, refined their single-epoch black hole mass estimates, and extended the $M_{\rm BH}$--$\sigma_*$ relation into the IMBH regime. High-resolution HST imaging of 147 objects showed that the hosts are predominantly disk-dominated systems with low S{\'e}rsic indices and weak or negligible bulges, demonstrating that IMBHs can reside in galaxies without classical bulges \citep{Jiang11}. In the X-rays, \citet{Dong12a} observed 49 sources with \textit{Chandra} and detected 42, with luminosities and spectral properties characteristic of AGN accretion. Their \textit{Spitzer} infrared spectra also resemble those of more massive AGNs and generally exhibit weak PAH emission \citep{Hood17}. At radio wavelengths, GH07 cross-matched their sample with the Faint Images of the Radio Sky at Twenty Centimeters (FIRST) survey and identified only 11 detections, consistent with the low radio-detection rate expected for low-mass AGNs. \citet{Yang22} subsequently observed the four radio-brightest sources with the milliarcsecond-resolution Very Long Baseline Array (VLBA), detecting compact nuclear radio emission consistent with AGN activity.

\subsection{The Dong et al. Sample}
Following the pioneering work of \citet{Greene07}, \citet{Dong12} expanded the sample to 309 sources in SDSS DR4 by improving spectral decomposition techniques, enabling the detection of low-Eddington-ratio, weak-broad-line AGNs. \citet{LiuH18} extended the methodology to DR7, constructing a sample of 513 objects, while \citet{LiuW25} applied the same approach to DR17, identifying 930 low-mass AGNs up to $z\sim0.6$. Together, these studies provide an increasingly complete census of active low-mass AGNs in the SDSS.

\citet{LiuH18} found that 102/513 ($\sim$20\%) sources have X-ray counterparts in ROSAT and 26 ($\sim$5\%) are detected by FIRST. Using deeper data from SKA pathfinders (e.g., LOFAR) together with a more thorough analysis of the FIRST images, \citet{Wu24} increased the radio detection rate to 151 sources ($\sim$30\%). Note, however, that this detection rate is likely influenced by several factors, including contamination from star formation and the limited spatial resolution of current radio surveys.

\subsection{The Chilingarian et al. Sample}
The sample of \citet{Chilingarian18} focuses on an even more extreme and challenging mass regime than previous IMBH samples, targeting black holes with masses below $2\times10^5\,M_\odot$. Based on SDSS DR7 spectra of nearly a million galaxies (at redshifts up to $z\sim0.3$), they employed a non-parametric emission-line fitting technique to detect broad H$\alpha$ components. After applying strict quality cuts, they identified 305 IMBH candidates, hosted by typically compact and low-luminosity galaxies. Cross-matching with archival X-ray and UV catalogs identified counterparts for only about 10 candidates ($\sim$3\%), providing additional evidence for accretion activity. In many cases, the broad-line component only becomes apparent after subtracting the host galaxy light, raising concerns that it may partly arise from modeling artifacts compared to other samples. This is further evidenced by their follow-up Magellan spectroscopy of 12 representative candidates, which detected potential broad H$\alpha$ in only 5 of them, while the remaining 7 showed no clear evidence of broad-line emission. These findings led the authors to estimate a contamination rate of $\sim$57\%, attributed to supernovae, transient stellar processes, or imperfect spectral modeling.

\subsection{Other Samples}
The above five works constitute the main body of our IMBH candidate sample. Other IMBH candidate samples have been assembled using diverse techniques. For example, UGC~06728 and NGC~4051 are benchmark low-mass Seyferts with reverberation-based black hole masses consistent with the IMBH regime \citep{Bentz16, Denney09, Fausnaugh17}. Additional candidates have been identified by selecting galaxies with strong optical or X-ray variability and then confirming the presence of broad lines, similar to IMBHs \citep{Ho16,Bernal25}. Since low-mass black holes are believed to reside in small and young dwarf galaxies with less growth, studies focusing on such specific populations have revealed hundreds of candidates, many of which exhibit broad lines and multiwavelength AGN signatures \citep{Reines13, Salehirad22, Mezcua24}. Given that the variability timescale of IMBHs is expected to be short (e.g., hours to days), scaling with black hole mass, \citet{Shin22} further monitored intra-night variability and proposed several of the best IMBH candidates in their sample based on rapid variability. The detailed selection criteria and methodologies of these works can be found in the original papers. Finally, studies that do not report their targets or broad-line properties are not included in our analysis.

\section{Uniform Reanalysis and Sample Characterization}\label{sec:results}
\subsection{Spectral decomposition}\label{sec:decomp}
Prior low-mass black hole samples were assembled under heterogeneous selection criteria, spectral decomposition prescriptions, and virial mass calibrations. To ensure uniformity across studies, we reanalyze all reported low-mass black holes with available SDSS spectroscopy (1360 objects), applying a single, uniform spectral-decomposition pipeline and recomputing black hole masses with a consistent calibrator. We therefore restrict our analysis to SDSS data, which provides homogeneous wavelength coverage, resolution, and flux calibration; non-SDSS spectra are inconsistently archived and more heterogeneous.

\begin{figure*}[!t]
    \centering
    \includegraphics[width=1\linewidth]{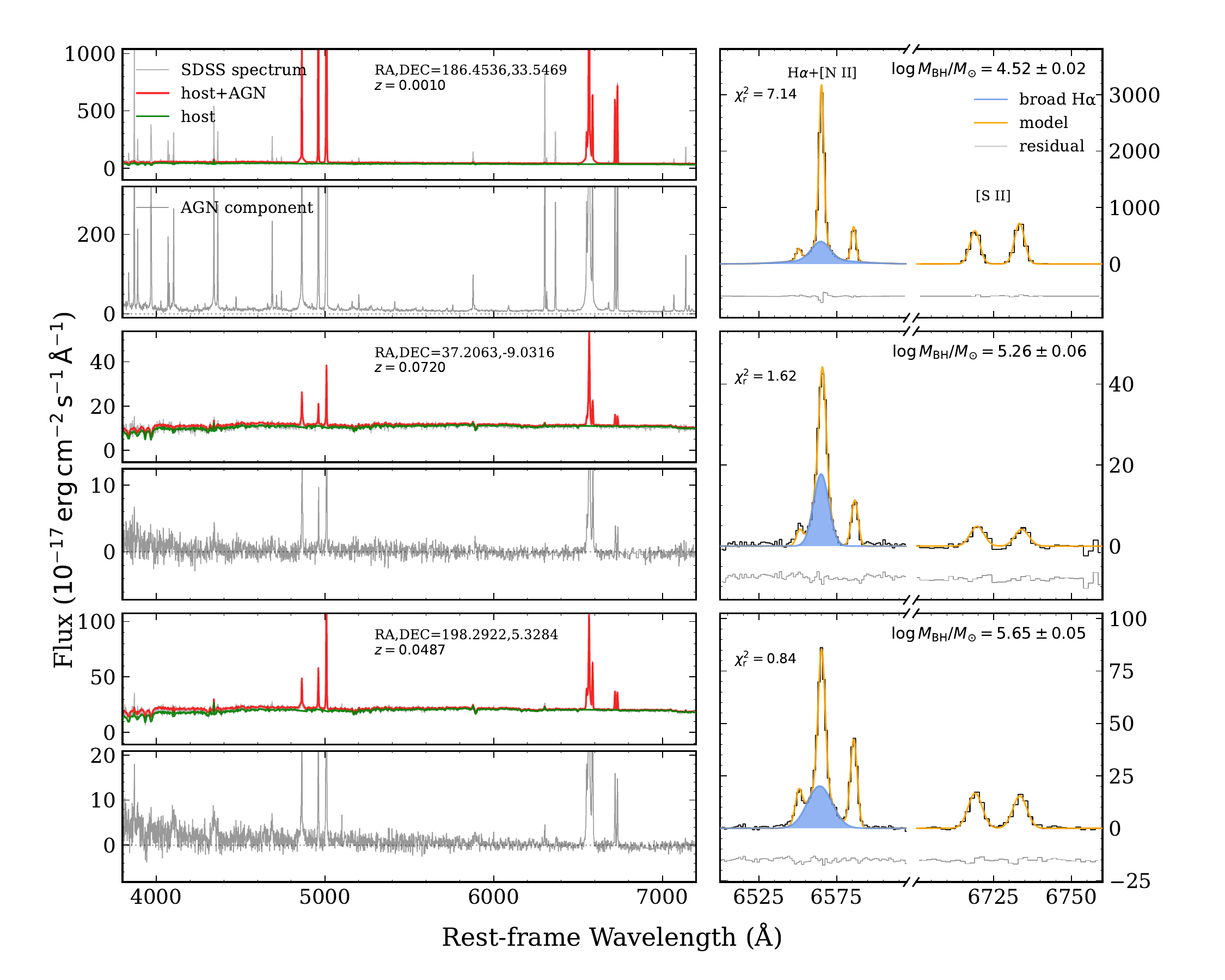}
    \caption{Three examples of the spectral decomposition. {\bf Left panels:} the upper panel shows the AGN--host separation, where the green curve represents the host-galaxy component derived from the PCA method, and the lower panel shows the host-subtracted AGN spectrum. {\bf Right panels:} enlarged views of the spectral decomposition in the H$\alpha$ region, with the corresponding reduced $\chi^2$ values indicated. The blue shaded region represents the broad H$\alpha$ component, used for black hole mass estimation. We emphasize that the PCA-based decomposition does not model the stellar H$\alpha$ absorption feature, thereby avoiding artificial broad H$\alpha$ emission produced by oversubtraction of the host-galaxy component. The spectral fits for all sources will be made available online.
    }
    \label{fig:fit_examples}
\end{figure*}

For IMBHs, broad emission lines are intrinsically weak and readily diluted by host starlight. Inadequate or excessive host subtraction can mimic weak broad features, requiring rigorous analysis. We perform spectral decomposition using \texttt{PyQSOFit} \citep{Guo18,Shen19}, a fitting code developed for quasar spectra and optimized for AGN component modeling. Figure \ref{fig:fit_examples} shows three representative examples spanning different black hole mass ranges. In the decomposition process, we utilized the prior-based PCA host modeling, which attains a $\sim$94\% success rate on SDSS DR16 quasars at $z<0.8$ \citep{Ren24}. The numbers of AGN and galaxy eigenspectra are set to 10 and 5, respectively, which are able to recover $\sim$98\%\ AGN spectrum \citep{Yip2004a,Yip2004b}. Regions containing broad lines are masked and interpolated linearly during fitting and host subtraction to prevent the creation of artificial features. After subtracting the host component, we model the residual AGN continuum using a power-law component and an optical Iron template.

While this procedure performs well for most of our sources, it may lack accuracy in cases dominated by host emission. In such spectra, strong galaxy absorption features can overwhelm potential weak broad emission from the AGN broad line region. However, the negligible AGN continuum contribution reduces the complexity of spectral decomposition, allowing for more precise modeling of galaxy absorption features. Consequently, sources identified with a host fraction at 5100 \AA\ exceeding 95\% in our initial decomposition are treated as pure galaxy spectra. We fit these directly using \texttt{pPXF} \citep{Cappellari17} after masking strong narrow lines. This program fits the data using synthetic stellar spectra with variable stellar velocity offsets and dispersion, providing stable and interpretable absorption estimates underlying the Balmer lines. We then subtract the best-fit galaxy model, leaving only the emission lines. Note that we correct for underlying stellar Balmer absorption in the pPXF decomposition, which is primarily applied to host-dominated spectra where stellar absorption could suppress part of the intrinsic broad emission. In contrast, the PCA decomposition is applicable mainly to spectra with relatively strong AGN emission, for which this correction is negligible; we therefore omit it to avoid introducing artificial broad-line components through imperfect host subtraction. Ultimately, 28.8\% (392) of the sources were decomposed using pPXF, while the remaining sources were decomposed using the PCA-based method implemented in PyQSOFit.

All emission lines are modeled with Gaussian profiles, using a single Gaussian for narrow lines and up to two Gaussians for broad components (Full Width at Half Maximum, FWHM $>500$~km~s$^{-1}$)\footnote{This choice follows \citet{Reines13} and represents a compromise between sample completeness and contamination, although it may exclude some of the lowest-mass black holes.} when necessary. The kinematics (velocity offsets and dispersions) of all narrow lines are tied, with line centers fixed by laboratory wavelength separations; the [N \textsc{ii}]$\lambda\lambda6548,6583$ flux ratio is fixed to the theoretical value of 2.96, while the [S \textsc{ii}]$\lambda\lambda6716,6731$ doublet is fitted with independent fluxes. As an exception, the [O \textsc{iii}]$\lambda\lambda4959,5007$ lines are modeled with a core+wing (two-Gaussian) profile to account for commonly observed outflows. The broad H$\beta$ component is modeled with a single Gaussian, as it is typically weak in low-mass systems. For the broad H$\alpha$ component, both single- and double-Gaussian models are tested, and the double-Gaussian fit is adopted only if it improves the reduced $\chi^{2}$ by at least 20\%, following the simple empirical criterion adopted in \citet{Dong12}. We adopt this conservative threshold to ensure consistency with previous IMBH studies and to avoid overfitting in moderate signal-to-noise (SNR) spectra, where information criteria may favor more complex models without providing physically meaningful improvements.

\begin{figure*}[!t]
    \centering
    \includegraphics[width=1\linewidth]{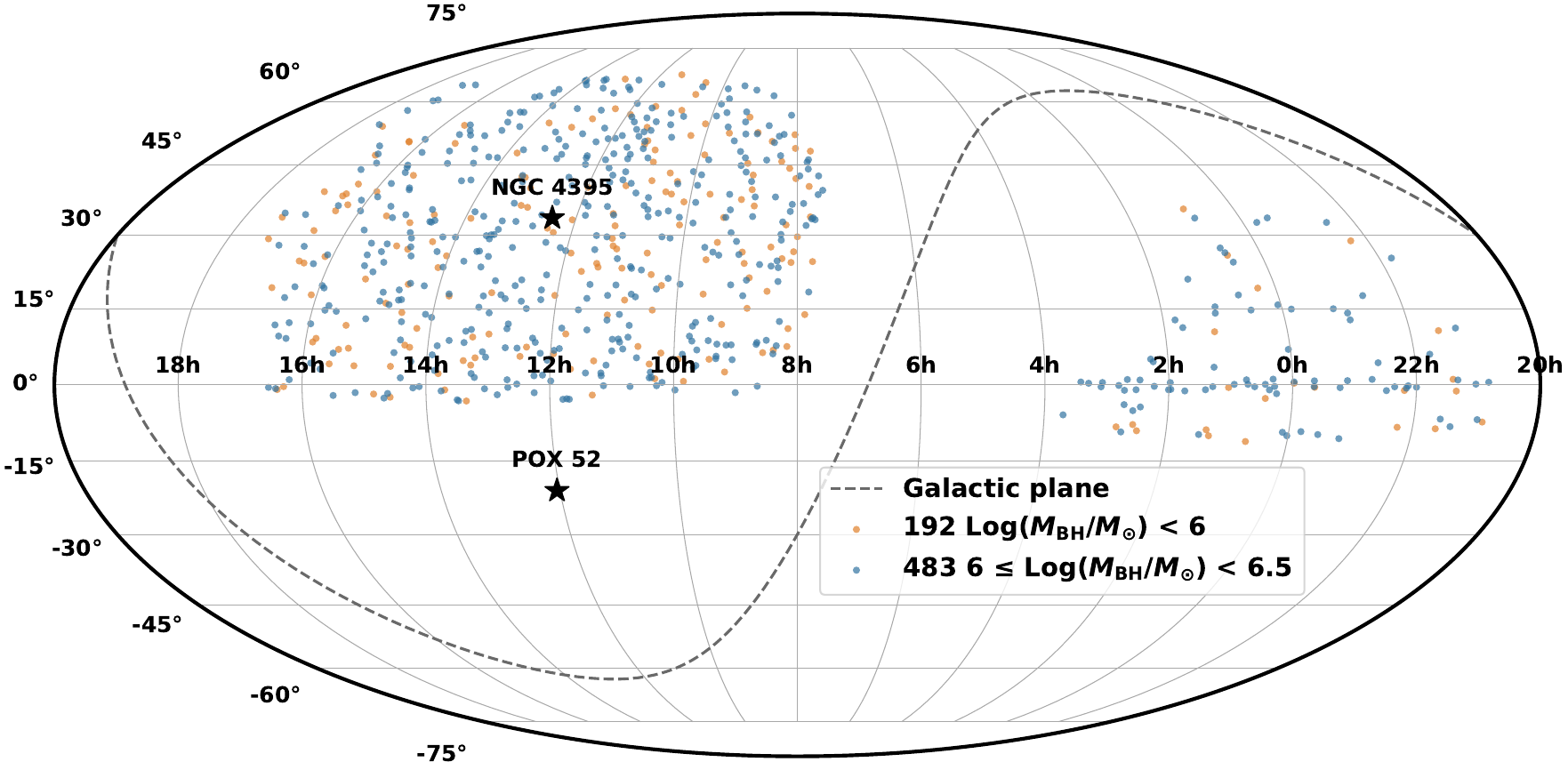}
    \caption{Sky distribution of the final low-mass AGN sample in equatorial coordinates. The source distribution closely follows the SDSS survey footprint, with only a small fraction located in the southern sky.}
    \label{fig:skymap}
\end{figure*}

\subsection{Robust broad line selection}
We next define quantitative criteria to identify broad emission-line components associated with robust IMBH activity. In practice, no unique set of conditions can cleanly separate IMBH-driven broad lines from stellar-related processes, nor from residuals arising from limitations of the line-profile modeling \citep[e.g., cases where Gauss–Hermite or Lorentzian profiles may provide a better description than simple Gaussians,][]{VeronCetty01,Riffel10}. Our workflow therefore begins with an initial visual inspection to assemble a high-confidence IMBH candidate sample, from which empirical quantitative boundaries are derived and iteratively refined to improve sample purity and reduce contamination from suspicious cases, followed by a final visual inspection for minor adjustments and overall quality control.

From these fits, we obtained the properties of the continuum and all emission-line components. To identify prominent broad H$\alpha$ emission associated with real IMBH activity, we apply the following empirical but physically motivated criteria to $\rm H\alpha$ line:
\begin{enumerate}
\item Broad $\rm H\alpha$ peak SNR $ > 3$;
\item F$({\rm broad\ H}\alpha)/\rm F({\rm continuum,\ RMS})
>80\,\mathring{\rm A}$\footnote{For a broad H$\alpha$ line with a peak SNR of 3, this threshold corresponds to a typical IMBH broad-line width of ${\rm FWHM}\sim1200\,{\rm km\,s^{-1}}$.};
\item F({\rm broad\ H}$\alpha$) $>$ 0.25\,F({\rm narrow\ H}$\alpha$);
\item Line width $\sigma_{\rm line}({\rm broad\ H}\alpha) > 2.5\sigma_{\rm line}({\rm narrow\ H}\alpha)$;
\item $c|\lambda_{\rm med}/6564.61\text{\AA} -1|/ \sigma_{\rm line}({\rm broad\ H}\alpha) < 0.7$;
\item If double-Gaussian fits, peak separation $<$ $\sigma_{\rm 1,line}$ + $\sigma_{\rm 2,line}$
\end{enumerate}

Criteria 1 and 2 jointly ensure the reliability of the broad H$\alpha$ component by requiring both a significant peak and sufficient integrated broad-line flux relative to the local continuum RMS, thereby rejecting spurious or marginal low-flux fits. Criterion 3 requires the broad component to contribute a non-negligible fraction ($\gtrsim$25\%) of the narrow component, ensuring a robust broad-line detection and increasing the likelihood of detecting line variability in subsequent RM campaigns. Criterion 4 enforces a clear kinematic separation between the broad and narrow components, reducing contamination from outflows, stellar-related processes, or model residuals. Criterion 5 constrains the centroid of the broad component relative to its width, excluding cases dominated by strong inflow/outflow signatures. Criterion 6 applies to double-Gaussian models and is designed to reject spurious broad H$\alpha$ detections caused by wing contamination from adjacent narrow [N \textsc{ii}] lines. 

Given that the adopted selection boundaries are empirical and that spectral decomposition can suffer from strong degeneracies, particularly when using simple Gaussian profiles that may not fully capture complex line shapes, we performed a visual inspection of all spectral fits by the five most experienced members of our team. Each classifier independently reviewed the entire sample, and objects were reassigned only if at least three out of the five classifiers agreed that a change in category was warranted. Through this process, 30 objects initially classified as high-confidence sources are reclassified as lower-significance candidates due to the weak or outflow like broad component, while 26 candidate objects showing convincing broad H$\alpha$ features are upgraded to the robust category, e.g., RGG 118 (see more discussion in Section \ref{sec:bias}). Based on these criteria, we assign a broad-line classification tag to all 1,360 candidates: 915 high-confidence broad-line sources are assigned $\mathrm{tag}_{\rm br}=1$, while the remaining 445 sources with relatively lower spectral SNR or less significant broad-line detections are assigned $\mathrm{tag}_{\rm br}=0$. We further exclude 240 of the 915 high-confidence candidates with $M_{\rm BH}\geq10^{6.5}\,M_\odot$. The final sample therefore consists of 192 robust IMBH candidates ($M_{\rm BH}<10^{6}\,M_\odot$; the core sample), together with 483 low-mass AGNs ($10^{6}$--$10^{6.5}\,M_\odot$) serving as a comparison sample. Their sky distribution is shown in Figure~\ref{fig:skymap}.

\begin{figure*}
  \centering
  \includegraphics[width=1.0\textwidth]{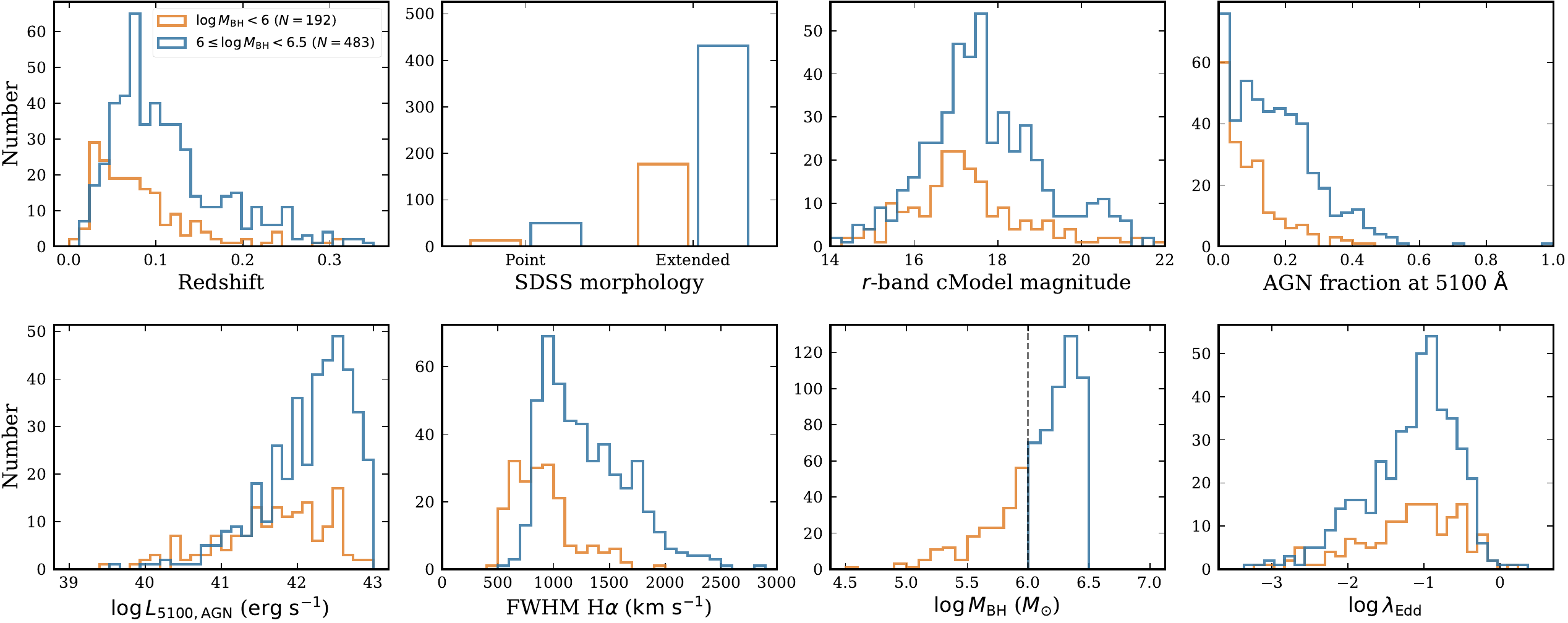}
    \caption{Distributions of the basic properties of our low-mass AGN sample. The panels show redshift, SDSS morphology, $r$-band cModel magnitude (uncertainties $<0.2$ mag), AGN fraction at 5100\,${\rm \AA}$, AGN continuum luminosity at 5100\,${\rm \AA}$, broad H$\alpha$ FWHM, black hole mass, and Eddington ratio. }
\label{fig:prop}
\end{figure*}

\subsection{Sample properties}
Figure~\ref{fig:prop} summarizes the other basic properties of the core and comparison samples. As expected, the IMBH core sample is preferentially found at lower redshift than the comparison sample, which hosts systematically more massive black holes. The vast majority of the IMBH candidates (177/192; 92.2\%) are classified as extended sources in the SDSS imaging, similar to the comparison sample (431/483; 89.2\%). Their median effective radius (Petrosian half-light radius) is $R_{\rm e}=2.18''$ ($\simeq2.65$ kpc), comparable to the comparison sample ($1.75''$, $\simeq2.99$ kpc).  The reported SDSS DR18 $r$-band cModel magnitudes include substantial host-galaxy light and therefore do not purely trace the nuclear emission. This host dilution is also evident in the AGN fraction at 5100\,\AA: the IMBH core sample is strongly host dominated, with a median $f_{\rm AGN}=0.068$, compared with 0.149 for the comparison sample. The fraction of sources with $f_{\rm AGN}>0.1$ is 37.5\% in the core sample, substantially lower than the 64.6\% found in the comparison sample. After accounting for host-galaxy contamination, the intrinsic AGN nuclei are likely even fainter, further reducing the number of suitable targets for future RM campaigns. Consistently, the core sample exhibits systematically lower AGN continuum luminosities at 5100~\AA\ and narrower broad H$\alpha$ lines, while having an Eddington-ratio distribution broadly similar to that of the comparison sample. Their Eddington ratios, mostly above 0.01, suggest that they are broadly accreting in the radiatively efficient, standard thin-disk regime.

All black hole masses are re-estimated using the single-epoch
virial calibration of \citet{Reines13},
\begin{equation}
\begin{aligned}
\log \left(\frac{M_{\rm BH}}{M_\odot}\right)
= log\epsilon+{} & 6.57
+ 0.47 \log \left(\frac{L_{{\rm H}\alpha}}{10^{42}\,{\rm erg\,s^{-1}}}\right) \\
& + 2.06 \log \left(\frac{{\rm FWHM}_{{\rm H}\alpha}}{10^3\,{\rm km\,s^{-1}}}\right),
\end{aligned}
\end{equation}
assuming an FWHM-based virial factor of $\epsilon=1$. As shown in Appendix~\ref{app:comparison}, our measurements for the core and comparison samples exhibit only small median offsets from the literature but substantial source-to-source scatter. This highlights the value of a homogeneous reanalysis, while recognizing that our adopted fitting procedure is not necessarily definitive.

\begin{figure*}[!tb]
    \centering
    \includegraphics[width=1\linewidth]{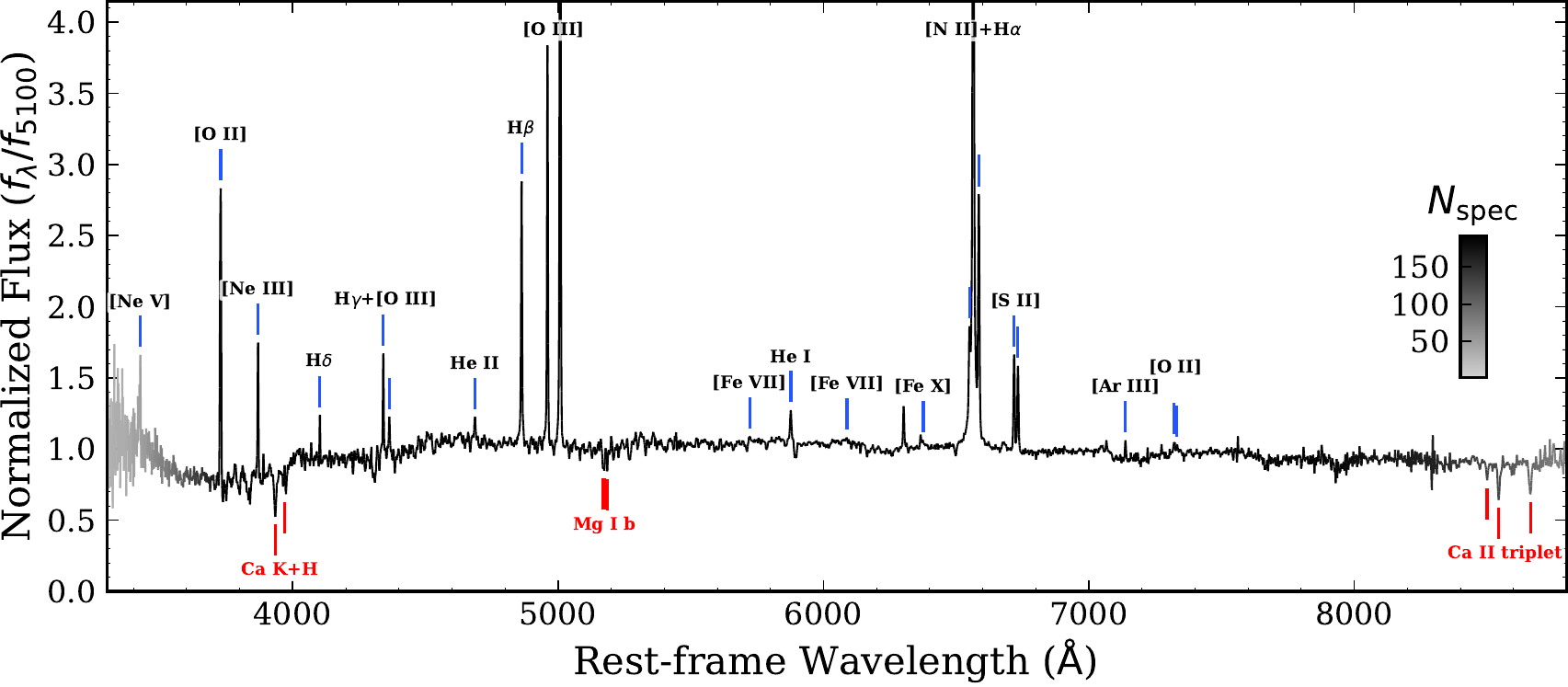}
    \caption{Composite spectrum of the 192 robust IMBH candidates with $\log(M_{\rm BH}/M_\odot)<6$, constructed using the arithmetic mean (data is available in Table \ref{tab:IMBH_composite}). Individual spectra are shifted to the rest frame and normalized by the continuum flux density at 5100~\AA\ before stacking. The grayscale indicates the number of spectra contributing at each wavelength. Major emission lines are labeled in black, while prominent stellar absorption features are marked in red. The 6370~\AA\ feature is a blend of [O I] $\lambda6364$ (dominated) and [Fe X] $\lambda6374$. The composite spectrum exhibits a relatively weak AGN continuum with clearly detected broad H$\alpha$ emission line.}
    \label{fig:composite}
\end{figure*}

\begin{figure}[!t]
    \centering
    \includegraphics[width=1\linewidth]{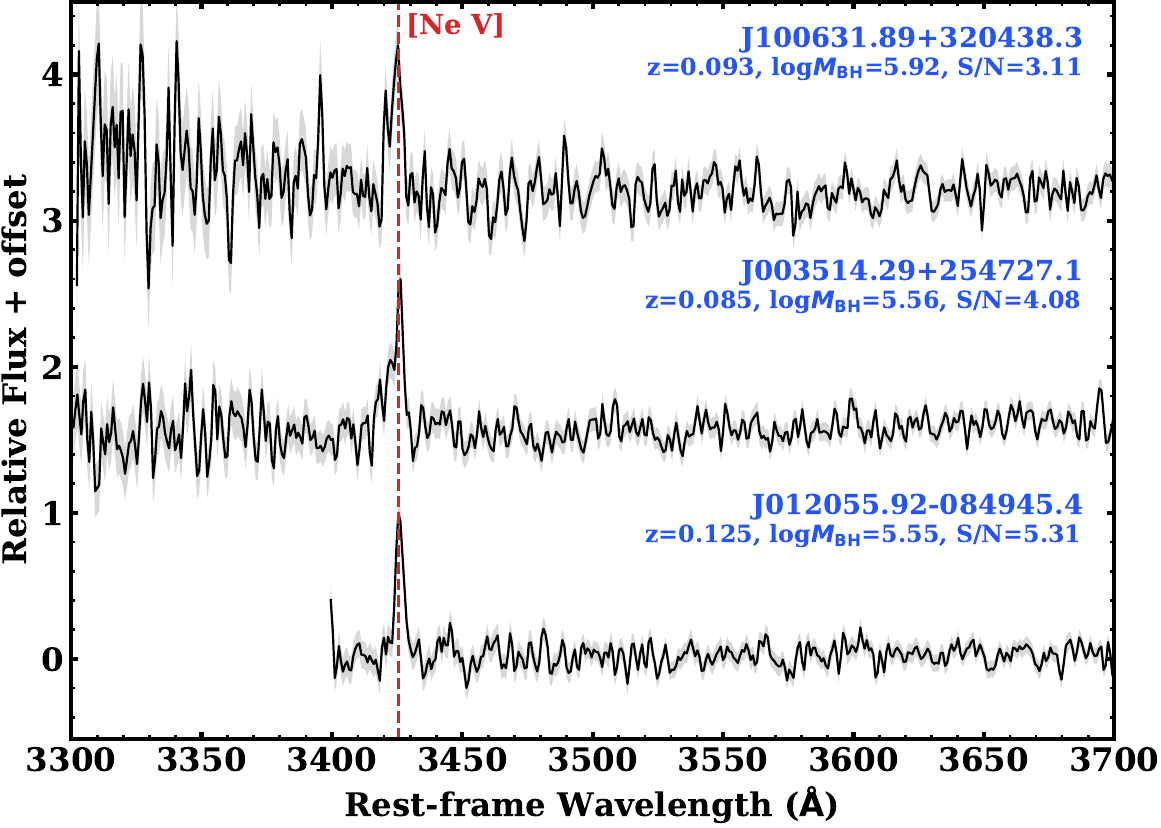}
    \caption{Three [Ne~V]-detected IMBH candidates with $\log(M_{\rm BH}/M_\odot)<6$. The spectra with $1\sigma$ uncertainties are vertically offset for clarity.}
    \label{fig:3NeV}
\end{figure}

\subsection{Composite spectrum}
To explore the overall spectral properties of 192 robust IMBH candidates, we construct a composite spectrum following \citet{Vanden01}. We adopt the arithmetic mean to maximally preserve the relative fluxes of emission-line features, with the composite flux density at each wavelength defined as $\langle F_\lambda \rangle = N^{-1}\sum_i F_{\lambda,i}$, where $F_{\lambda,i}$ is the rest-frame flux density of the $i$th spectrum and $N$ is the number of spectra contributing at that wavelength. As shown in Figure~\ref{fig:composite}, the composite spectrum exhibits a relatively flat continuum and generally weak broad emission lines compared with the typical AGN composite \citep{Vanden01}, suggesting that the weak AGN emission is strongly diluted by host-galaxy starlight rather than reflecting intrinsic evolution of the AGN spectrum \citep{Cai23}. The weak 4000~\AA\ break indicates a relatively young stellar population with ongoing or recent star formation in the host galaxies. A prominent broad H$\alpha$ component is clearly visible, while the broad H$\beta$ component is slightly less conspicuous. Moderately ionized emission lines, including [\ion{Ne}{3}] $\lambda3869$ (ionization potential 40.96 eV), He~I (24.59 eV), He~II $\lambda4686$ (54.42 eV), [\ion{Ar}{3}] $\lambda7136$ (27.63 eV), are clearly detected. In contrast, higher-ionization coronal lines are generally absent (e.g., [\ion{Fe}{7}] and [\ion{Fe}{10}]), while the [\ion{Ne}{5}] $\lambda\lambda3425,3426$ doublet (97.1 eV) is only marginally visible in the composite spectrum. We therefore systematically inspected all sources with $\log(M_{\rm BH}/M_\odot)<6$ and identified significant ($>3\sigma$) [\ion{Ne}{5}] emission in only three core-sample objects, shown in Figure~\ref{fig:3NeV}, providing clear evidence of AGN activity. No significant [\ion{Ne}{5}] emission is detected among candidates with less secure broad-line detections ($\mathrm{tag}_{\rm br}=0$). All three detections have black hole masses in the range $\log(M_{\rm BH}/M_\odot)\simeq5.5$--6. After excluding sources whose spectra do not cover [\ion{Ne}{5}], the detection rate is 3/91 $\sim$ 3.3\% with $\log(M_{\rm BH}/M_\odot)<6$. This incidence is broadly consistent with the detection rate of 4.4\% reported by \citet{Doan25} for 19,508 SDSS Type~1 quasars at $z<0.8$, given our sample probing substantially lower black hole masses.

\begin{table}
\centering
\caption{Mean composite spectrum for the $\log(M_{\rm BH}/M_\odot)<6$ IMBH sample. The complete table is available online.}
\label{tab:IMBH_composite}
\begin{tabular}{ccc}
\hline
Wavelength ($\mathrm{\AA}$) & Mean normalized flux & $N_{\rm spec}$ \\
\hline
3000.0 & 1.254898 & 9 \\
3001.0 & 1.720106 & 9 \\
3002.0 & 0.970207 & 9 \\
3003.0 & 1.491011 & 9 \\
3004.0 & 1.404439 & 9 \\
\hline
\end{tabular}
\end{table}

In addition, following \citet{Xiao11}, we measure the stellar velocity dispersion from the high-SNR Ca II triplet, whose strong absorption features are less affected by AGN continuum dilution and stellar population variations than blue stellar absorption lines. Fitting the stacked spectrum with pPXF yields $\sigma_{\rm obs}=104.7\pm9.2~{\rm km\,s^{-1}}$, corresponding to an intrinsic dispersion of $\sigma_*=78.7\pm12.2~{\rm km\,s^{-1}}$ after correcting for the SDSS instrumental resolution of 69 km s$^{-1}$. The active-galaxy $M_{\rm BH}$--$\sigma_*$ relation then implies a nominal black hole mass scale of $\log(M_{\rm BH}/M_\odot)\approx6.3$, with an intrinsic scatter of $\sim$0.46 dex \citep{Xiao11}. Because residual broadening may arise from spectral stacking, imperfect rest-frame alignment, and source-to-source velocity offsets, this estimate should be regarded as an approximate upper limit. Nevertheless, it is broadly consistent with the characteristic single-epoch virial black hole masses of our sample.

We further examine composite spectra constructed from the comparison sample with secure broad-line detections and find that they exhibit similar overall spectral properties, but with slightly broader Balmer emission lines. In contrast, the composite spectrum of the low-confidence IMBH candidates ($\mathrm{tag}_{\rm br}=0$) shows that the broad Balmer components are largely absent. We also construct geometric-mean composite spectra and find no significant differences from the arithmetic-mean results.

\begin{figure}[h]
    \centering
    \includegraphics[width=1\linewidth]{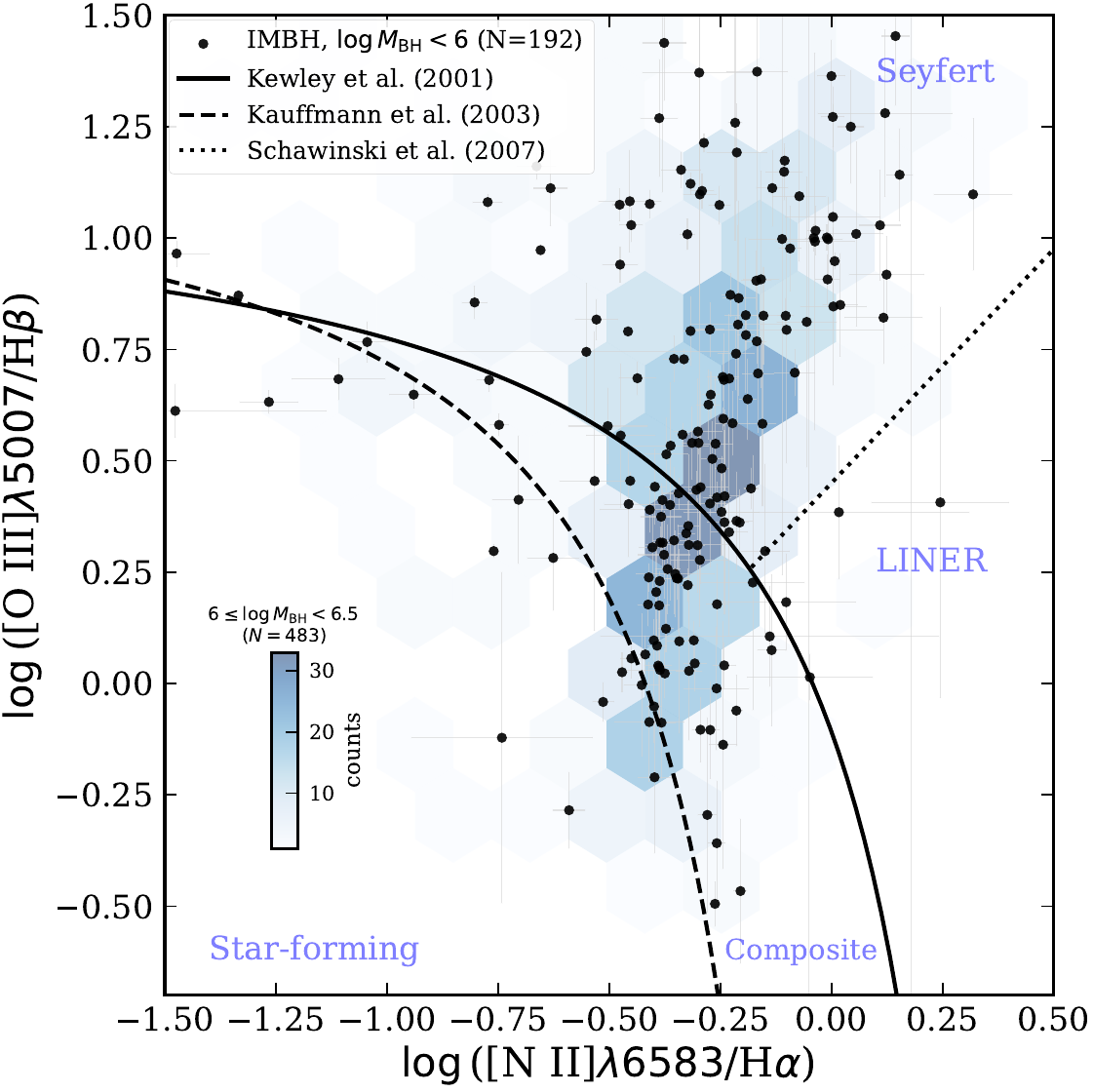}
    \caption{The [N II]-based BPT diagnostic diagram for the secure IMBH candidates. Black circles represent sources with $\mathrm{tag}_{\rm br}=1$ and \(\log(M_{\rm BH}/M_\odot)<6\), with gray error bars propagated from the narrow-emission-line flux uncertainties. The blue hexagonal distribution shows the number density of secure candidates with \(6\le\log(M_{\rm BH}/M_\odot)<6.5\), with the counts per bin indicated by the color bar. }
    \label{fig:bpt}
\end{figure}

\subsection{BPT diagram}
To assess whether the identified broad H$\alpha$ components are associated with nuclear accretion, we examine the [N~II]-BPT diagram \citep{Baldwin81}. Among the 192 robust IMBH candidates (core sample, $\log(M_{\rm BH}/M_\odot)<6$), 112 (58.3\%) lie in the Seyfert region, 60 (31.3\%) in the composite region, and only 16 (8.3\%) and 4 (2.1\%) occupy the star-forming and LINER regions, respectively. We visually inspected the spectra of all 20 objects in the latter two regions and found no obvious problems with their broad H$\alpha$ detections or spectral fits. The BPT star-forming objects are not preferentially lower-luminosity broad-line sources, with a median $\log (L_{\rm H\alpha,broad}/\rm erg\, s^{-1})=40.45$, comparable to the rest of the core sample. Thus, there is no clear evidence that their broad H$\alpha$ emission preferentially originates from stellar-related processes, although contributions from massive evolved stars, stellar winds, supernovae, or supernova remnants cannot be ruled out for individual objects \citep[e.g.,][]{Filippenko97}. A comparison sample with $6\leq\log(M_{\rm BH}/M_\odot)<6.5$ shows a similar BPT distribution. Both samples are therefore dominated by Seyfert and composite systems, indicating that AGN photoionization contributes significantly to the narrow-line emission in most sources.

The core sample has median line ratios of
$N2=\log([\mathrm{N\,II}]/\mathrm{H}\alpha)=-0.297$ and
$O3=\log([\mathrm{O\,III}]/\mathrm{H}\beta)=0.581$. Compared with
the Seyfert population in the SDSS DR4 AGN catalog
\citep{Kauffmann03}, the core sample has a similar median $O3$ value
(0.581 versus 0.573) but a lower median $N2$ value by approximately
0.19 dex. This behavior is consistent with previous findings that
low-mass AGNs extend to systematically lower
[\ion{N}{2}]/H$\alpha$ ratios than typical Seyferts
\citep{Ludwig12}. Although lower $N2$ values may qualitatively reflect
lower narrow-line-region gas-phase metallicities, $N2$ also depends on
the ionization parameter, the N/O abundance ratio, aperture effects,
and sample selection. The observed offset therefore cannot be
uniquely attributed to metallicity \citep{Groves06,Carvalho20}.

\begin{figure*}
    \centering
    \includegraphics[width=1\linewidth]{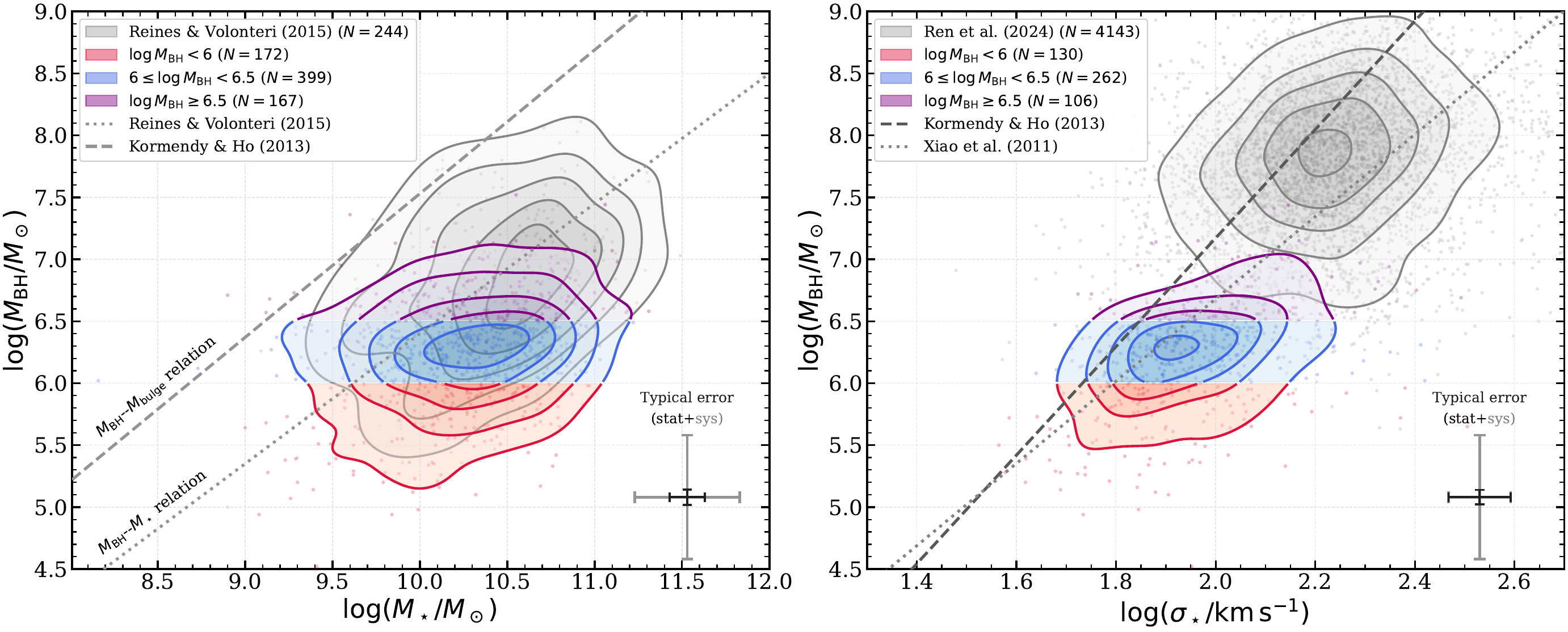}
    \caption{Black hole--host scaling relations. {\bf Left panel}: $M_{\rm BH}$ versus total stellar mass $M_\star$.  The KH13 relation shows the canonical local $M_{\rm BH}$--$M_{\rm bulge}$ relation for dynamically measured black holes in classical bulges and elliptical galaxies, while the R15 relation is the local $M_{\rm BH}$--$M_\star$ relation for broad-line AGNs based on single-epoch virial black hole masses and total stellar masses. {\bf Right panel}: $M_{\rm BH}$ versus stellar velocity dispersion $\sigma_*$, compared with the broad-line AGN sample of \citet{Ren24}. Both samples include only galaxies with $\sigma_\star$ measured at $>3\sigma$; the \citet{Ren24} sample further requires host-galaxy S/N $>4$ and $\sigma_\star<500~{\rm km~s^{-1}}$, yielding 4,143 AGNs. Contours show two-dimensional Gaussian KDEs enclosing 90, 75, 55, 35, and 15\% of the distribution, from outermost to innermost.}
    \label{fig:relations}
\end{figure*}

\subsection{$M_{\rm BH}$--$M_\star$ relation}

Of the 192 robust IMBH candidates, 172 have stellar-mass measurements, with a median $\log(M_\star/M_\odot)=10.16\pm0.47$ from heterogeneous measurements\footnote{Stellar masses are compiled from four catalogs following a hierarchical priority scheme based on SED wavelength coverage and modeling capability: good-quality fits with broader wavelength baselines are prioritized, adopting the DESI DR1 VAC \citep[UV-to-mid-IR with CIGALE, which additionally models the AGN component; requiring $1 < \chi^2 < 10$,][]{Siudek24}, GSWLC-2 \citep[UV-to-mid-IR,][]{Salim16}, and  \citep[optical-to-mid-IR; $\mathrm{flag}=1$,][]{Chang15} in order, followed by lower-quality SED fits, and finally optical-only masses from the NASA-Sloan Atlas (\url{https://www.nsatlas.org/}) as fallbacks.}. Only 19 objects (11.1\%) have $\log(M_\star/M_\odot)<9.5$, indicating that most are not hosted by typical dwarf galaxies. HST imaging studies of the GH07 sample, which provides an important and representative subset of our sample, found that 93\% of the hosts contain extended disks and that most host galaxies harbor pseudobulges rather than classical bulges \citep{Greene08,Jiang11}.

Figure~\ref{fig:relations} (left panel) shows that our sample extends the broad-line AGN distribution of \citet{Reines15} toward lower black hole masses at comparable total stellar masses, while lying significantly below the classical $M_{\rm BH}$--$M_{\rm bulge}$ relation of \citet{Kormendy13} (KH13). Dynamically measured black holes in classical bulges typically have $M_{\rm BH}/M_{\rm bulge}$ ratios of order of $10^{-3}$ \citep{Haring04,Kormendy13}. For the 738 broad-line AGNs in our sample with valid stellar-mass measurements, the median $M_{\rm BH}/M_*$ ratio is $10^{-4}$, about a factor of 2.5 lower than the characteristic value of $2.5\times10^{-4}$ reported for the broad-line AGNs of \citet{Reines15}. Restricting to the 172 robust IMBH candidates with $\log(M_{\rm BH}/M_\odot)<6$, the median $M_{\rm BH}/M_*$ ratio is $3.7\times10^{-5}$. Relative to the broad-line AGN relation in \citet{Reines15}, this is lower by a factor of $\sim6$--$7$. Thus, our active IMBH candidates appear modestly undermassive relative to their host galaxies compared with typical low-redshift broad-line AGNs in \citet{Reines15}. However, this interpretation is subject to uncertainties in virial black hole masses and stellar-mass estimates. In particular, except for the DESI CIGALE catalog, the adopted stellar masses are derived without explicitly accounting for AGN contribution during SED fitting. While attributing unmodeled AGN light to host starlight tends to inflate the total luminosity, the blue AGN continuum can simultaneously bias fitted stellar populations toward younger ages and lower mass-to-light ratios. These competing effects introduce additional systematic uncertainties into $M_\star$ and the inferred $M_{\rm BH}/M_\star$ offsets. Our sample is also subject to selection effects (see details in \S \ref{sec:bias}).

The decreasing $M_{\rm BH}/M_*$ ratio may reflect both changing host structure and different stages of black hole--galaxy coevolution toward lower black hole masses. The KH13 relation uses bulge stellar mass and is dominated by classical bulges and ellipticals, whereas \citet{Reines15} and our work use total stellar mass for broad-line AGNs. HST decompositions of the GH07 sample show that 93\% of the hosts contain extended disks, 76/147 are disk dominated with $B/T<0.2$, and seven have no detected bulge component \citep{Jiang11}. Our lower-mass IMBH candidates may therefore include a larger fraction of disk-dominated and pseudobulge hosts, naturally reducing $M_{\rm BH}/M_*$. Delayed black hole growth may also contribute, as stellar feedback in low-mass galaxies can suppress sustained accretion and allow the host to grow before the black hole approaches the local scaling relation \citep{AnglesAlcazar17,Byrne23}. These systematic changes in host structure and evolutionary stage are why we do not fit all datasets with a single linear relation. Nevertheless, the observed $M_{\rm BH}$--$M_*$ distribution shows no clear flattening at the low-mass end, consistent with a light-seed scenario \citep[see also][]{Pucha25,Pucha26} and in contrast to the elevated $M_{\rm BH}/M_*$ ratios reported for higher-redshift AGNs \citep[e.g.,][]{Ding20,Harikane23,Maiolino24}.

\subsection{M-$\sigma_*$ relation}
The $M_{\rm BH}$--$\sigma_\star$ relation is often regarded as more physically fundamental than the $M_{\rm BH}$--$M_\star$ relation because stellar velocity dispersion more directly traces the central gravitational potential. As emphasized by KH13, $\sigma_\star$ is also less sensitive to galaxy assembly history, stellar-population assumptions, and mass-to-light ratios, and therefore provides a more intrinsic probe of the connection between black holes and their host galaxies. 

To construct this relation in Figure~\ref{fig:relations} (right panel), we include only galaxies with $\sigma_\star>3\sigma$ in both our sample and that of \citet{Ren24}. We further assess the quality of our SDSS-based measurements in Appendix~\ref{app:AppB} by comparing them with the higher-resolution measurements of \citet{Xiao11}. For 63 matched sources, the median offset is only 0.002 dex, although the SDSS measurements have larger uncertainties. Two caveats should be noted. First, the SDSS instrumental resolution of $\sim69~{\rm km~s^{-1}}$ limits the reliability of measurements near or below this value. Second, unresolved disk rotation may artificially increase $\sigma_\star$, particularly in inclined disk galaxies, shifting sources toward larger $\sigma_\star$ \citep{Xiao11}.

Our sample forms a continuous extension of the broad-line AGN $M_{\rm BH}$--$\sigma_\star$ relation toward lower black hole masses and is broadly consistent with both the low-mass AGN relation of \citet{Xiao11} and the extrapolation of the classical-bulge relation of KH13. Most importantly, no clear turnover or flattening is seen down to $M_{\rm BH}\sim10^{5}\,M_\odot$, consistent with the behavior of the $M_{\rm BH}$--$M_\star$ relation. Nevertheless, scaling relations calibrated using samples with measurable black hole masses may be affected by non-representative host selection, as demonstrated at the high-mass end \citep{Lauer07}. This smooth continuation does not support the strong low-mass flattening expected in some heavy-seed scenarios and is more consistent with a substantial contribution from light black hole seeds \citep[e.g.,][]{Mezcua17,Greene20,Volonteri21}. However, the absence of such a feature does not uniquely favor light seeds, because subsequent accretion and galaxy evolution can erase the initial seeding signatures, making the low-mass end of the local $M_{\rm BH}$--$\sigma_\star$ relation a poor discriminator between seeding scenarios \citep{Ricarte18}. 

\begin{figure}[h]
    \centering
    \includegraphics[width=1\linewidth]{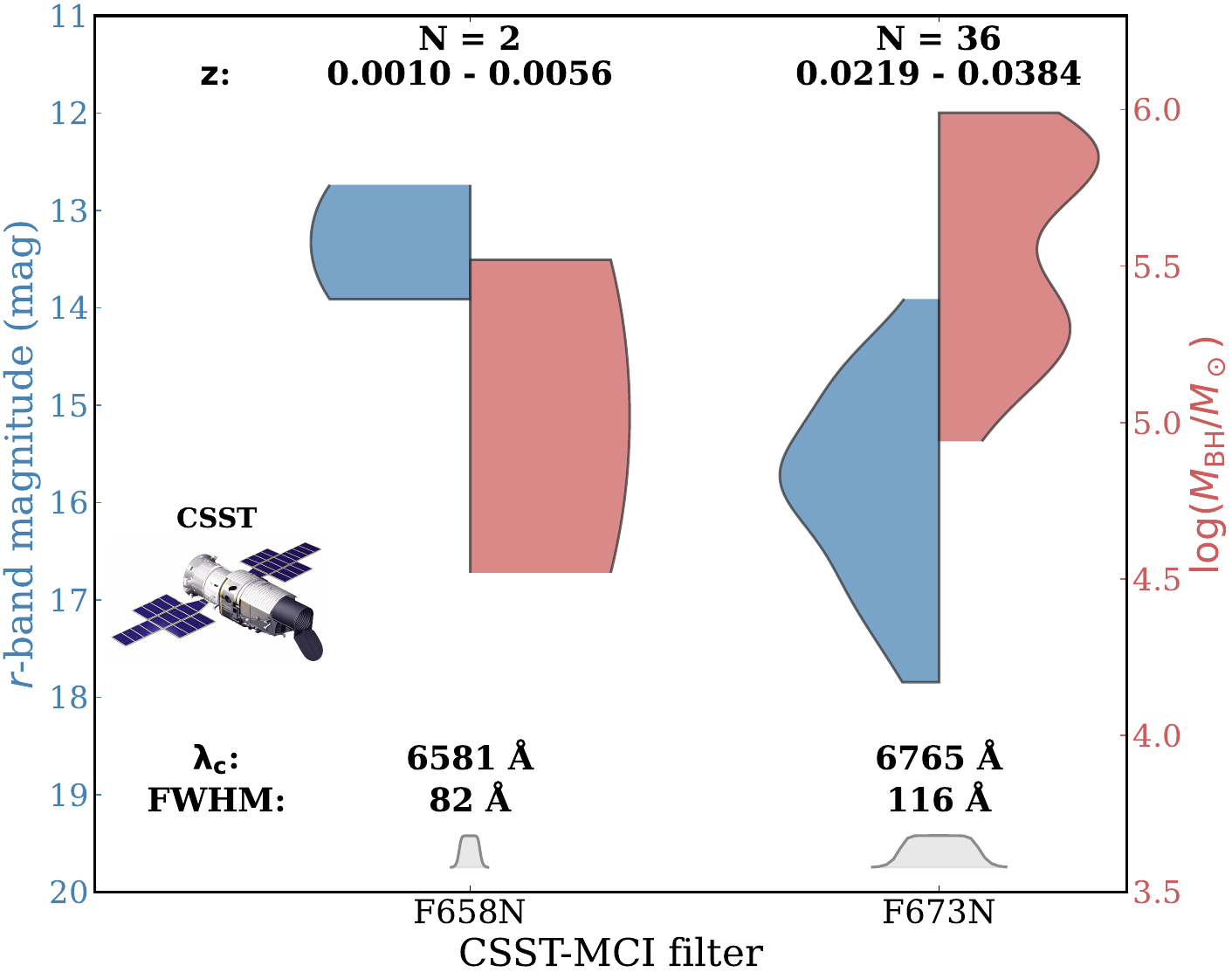}
    \caption{Distributions of observed $r$-band cModel magnitude (blue; left axis) and virial black hole mass (red; right axis) for the 38 robust IMBH candidates with $\log(M_{\rm BH}/M_\odot)<6$ and $r<18$ mag (see also Table \ref{tab:mci}). Gray curves show the original MCI filter transmission profiles. }
    \label{fig:mci}
\end{figure}

\subsection{Photometric RM with CSST/MCI}
CSST’s Multi-Channel Imager \citep[MCI;][]{Zheng25} provides simultaneous three-channel imaging over 2550–10000~\AA, enabling stable time-series photometry with minute-level cadence. Its greatest advantages for IMBH RM are the absence of atmospheric seeing and transparency variations, together with the fine 0.05 arcsec pixel$^{-1}$ image sampling, enabling high-precision measurements of weak nuclear variability with minimal host-galaxy contamination. These capabilities are particularly well suited for detecting the rapid continuum and broad-line variability expected in IMBHs. The numerous narrow band filters make MCI ideally suited for photometric RM. Although the CSST integral-field spectrograph (IFS) is also capable of spectroscopic RM, such observations are considerably more expensive.

To identify the proper IMBH targets for future RM campaigns, we select robust candidates with $\log(M_{\rm BH}/M_\odot)<6$ and $r<18$ mag, requiring at least 50\% of the modeled broad H$\alpha$ profile (assuming $\lambda_{\rm H\alpha}=6564.6$ \AA\ in vacuum) to fall within the 50\%-transmission interval of an MCI narrow band filter. Note that the listed r-band magnitudes represent the total galaxy light; the AGN nuclei themselves are typically substantially fainter. This yields 38 prime targets shown in Figure~\ref{fig:mci}. On average, broad H$\alpha$ contributes $f_{\rm broad}^{\rm line}\sim 0.4$ of the transmitted emission-line flux and $f_{\rm broad}^{\rm filter}\sim0.1$ of the total narrow band flux. The individual values are listed in Table~\ref{tab:mci} for future exposure-time and lag-recovery simulations. Based on their AGN continuum luminosities at 5100 \AA, the predicted BLR lags are generally $\lesssim2.5$ days, while the interband ($g-z$) continuum lags are expected to be $<5$ hrs, following the BLR and continuum $R-L$ relations in \citet{Wang23}. A few additional candidates are covered by the F815N filter but are likely too faint for efficient high-cadence monitoring. Looking further ahead, the proposed HWO telescope, with its much higher sensitivity and angular resolution, could extend IMBH RM to substantially fainter nuclei \citep{Feinberg26}.

\subsection{Cross-match with X-ray and radio observations}

\subsubsection{X-ray Observations}

The detection of sufficiently high X-ray luminosities in galaxies is commonly regarded as compelling evidence for accreting black hole activity \citep{Brandt15}. We cross-matched our whole sample with catalogs from three major X-ray observatories: the Chandra Source Catalog Version 2.1 \citep[CSC 2.1;][]{2024ApJS..274...22E}, the \textit{XMM-Newton} source catalogs including the serendipitous catalog 4XMM-DR14 \citep{2020A&A...641A.136W}, the stacked catalog 4XMM-DR14s \citep{Traulsen_2020}, and the slew survey catalog XMMSL3 \citep{Saxton_2008}, as well as the eROSITA All-Sky Survey Data Release 2 \citep[eRASS-DR2;][]{eROSITA_DR2_26}. In addition, we reprocessed archival \textit{Chandra} and \textit{XMM-Newton} observations not included in these catalogs in order to update the X-ray information to January~1,~2026.

\textit{Chandra} and \textit{XMM-Newton} are pointed X-ray observatories with complementary capabilities: \textit{Chandra} provides arcsecond-level angular resolution (0.5--7 keV) for accurate localization of nuclear emission, while \textit{XMM-Newton} offers a larger effective area and broad 0.2--12 keV coverage for sensitive detection of faint sources. In contrast, eROSITA is designed as a uniform all-sky survey; however, the current eRASS-DR2 release contains only the German half of the survey (Galactic longitudes $180^\circ$--$360^\circ$), reaching a typical depth of $\sim2.7\times10^{-14}$ erg s$^{-1}$ cm$^{-2}$ in the 0.5--2.0 keV band. For the \textit{Chandra} data, we adopted a nearest-neighbor positional matching radius of $2''$ with 1447 low-mass black holes, yielding 161 X-ray detections. Owing to the larger point-spread function of \textit{XMM-Newton}, a $10''$ matching radius was used, resulting in 169 detections from the combined \textit{XMM-Newton} catalogs. We also cross-matched our sample with the eRASS-DR2 catalog using a $10''$ matching radius, identifying 270 counterparts among the 578 sources located within the German eROSITA survey footprint, corresponding to a detection fraction of $46.7\%$.

\begin{figure*}[ht]
    \centering
  \includegraphics[width=1\linewidth]{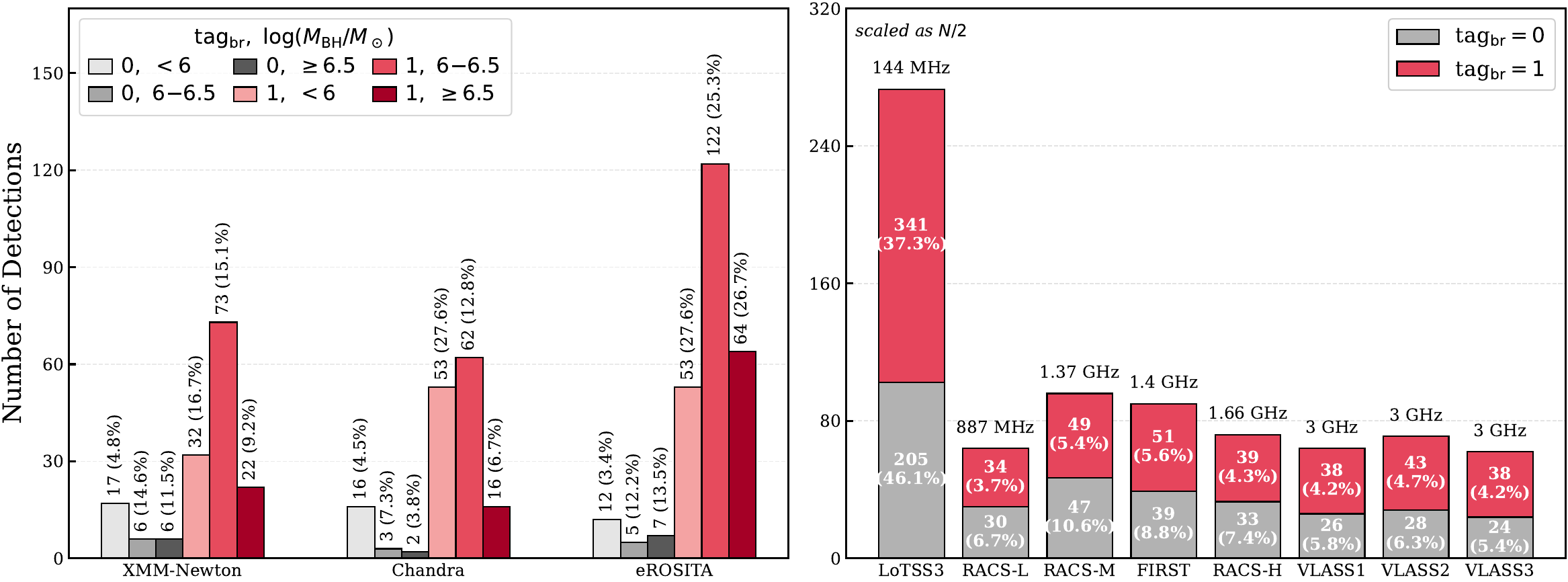}
    \caption{Detection statistics for the 1360 sources with valid black hole mass estimates, comprising 915 sources with $\mathrm{tag}_{\rm br}=1$ and 445 with $\mathrm{tag}_{\rm br}=0$. \textbf{Left panel:} X-ray detections grouped by X-ray survey, broad-line tag, and black hole mass bin. \textbf{Right panel:} Radio detections grouped by radio survey and broad-line tag. Bars indicate the number of detected sources, with percentages calculated relative to the total number of sources in the corresponding $\mathrm{tag}_{\rm br}$ w/o $\log(M_{\rm BH}/M_\odot)$ bin. For clarity, the LoTSS3 bars are scaled by a factor of 1/2.}
    \label{fig:xray_radio}
\end{figure*}

Figure~\ref{fig:xray_radio} (left panel) and Table~\ref{tab:xray} summarize the X-ray detection statistics for the 1360 sources with valid black hole mass estimates. After removing duplicates, 454 of these sources are detected in at least one X-ray survey, corresponding to an overall detection rate of 33.4\%. The main trend of Figure~\ref{fig:xray_radio} is that sources with higher confidence broad-line components have both a larger absolute number of X-ray detections and a higher detection fraction than those with lower confidence broad lines, especially in the homogeneous eROSITA survey. However, an X-ray detection alone neither provides definitive evidence for AGN activity nor confirms that the identified broad-line component is genuine, since X-ray emission can also arise from star formation and X-ray binaries, particularly at lower luminosities \citep[e.g.,][]{Mineo12}. We therefore further examine the luminous X-ray sources ($L_{\rm X}>10^{42}$ erg s$^{-1}$), which are more likely to be powered by AGN activity. For sources with $\log(M_{\rm BH}/M_\odot)<6$, 58/100 (58.0\%) of the X-ray-detected sources with $\mathrm{tag}_{\rm br}=1$ have $L_{\rm X}>10^{42}$ erg s$^{-1}$, compared with 11/35 (31.4\%) of the sources with $\mathrm{tag}_{\rm br}=0$, providing additional statistical support for the broad-line sample. In addition to the 1360 sources with valid black hole mass estimates, 19 of the remaining 87 sources are also detected in X-rays. A more detailed analysis based on hard X-ray detections, X-ray spectral properties, and other diagnostics of our sample will be presented in Tian et al. in prep..

\subsubsection{Radio properties}
To characterize the radio properties of our sample across a broad frequency range, we cross-matched the sources with four major large-area radio surveys, spanning from low-frequency (144 MHz) to GHz bands. Specifically, we used LoTSS DR3 \citep{Shimwell26} at 144 MHz to probe low-frequency radio emission with high sensitivity, FIRST \citep{Becker95} at 1.4 GHz and VLASS \citep{Lacy20} at 3 GHz to characterize GHz-band radio cores with progressively higher angular resolution, and the RACS survey \citep{McConnell20} at 0.88 to 1.65 GHz to provide complementary coverage of the southern sky at intermediate frequencies.

For all surveys, we performed a positional cross-match by identifying the closest radio source within a fixed search radius. Figure~\ref{fig:xray_radio} (right panel) and Table~\ref{tab:radio_surveys} summarize the results for the 1360 SDSS sources. Unlike the X-ray results, sources with $\mathrm{tag}_{\rm br}=1$ have higher radio number detections but consistently lower detection fractions than the lower confidence sample ($\mathrm{tag}_{\rm br}=0$) across all radio surveys. This suggests that the radio counterparts of the lower confidence sample are more likely to arise from star formation or other non-AGN processes \citep{Wu24}. This is also consistent with their BPT classifications, where most lie in the star-forming region. Therefore, we do not further divide the sample into different mass bins in the following discussion. Furthermore, arcsecond-scale radio emission does not necessarily trace the currently accreting nucleus. For example, in NGC~4395, one of the most robust IMBHs, milliarcsecond-resolution radio observations suggest that the detected radio emission is associated with shocks produced by an outflow interacting with the surrounding interstellar medium, rather than with compact nuclear emission from the central black hole \citep{Yang22}. Therefore, confirming whether these sources host genuine nuclear radio emission now will require future high-resolution high-sensitivity observations, for example with VLBI capabilities provided by the ngVLA \citep{Wilner24} and the SKA \citep{Paragi15}.

\section{Discussion}\label{sec:diss}

\subsection{Selection Bias}\label{sec:bias}
Our sample is designed to provide a high-purity set of broad-line IMBH candidates suitable for future RM campaigns, rather than a complete census of the IMBH population. This choice introduces several selection biases. First, by requiring detectable broad Balmer emission, we are necessarily biased toward unobscured, actively accreting type I nuclei, and miss inactive black holes, obscured AGNs, and systems in which the broad-line region is intrinsically weak or absent. Second, our conservative broad H$\alpha$ criteria favor sources with high-contrast, well-separated broad components. These thresholds are empirically defined to improve sample purity, rather than physically sharp divisions. They are essential for suppressing contamination from outflows, stellar processes, imperfect narrow-line modeling, and host-subtraction residuals, but they also bias the sample against low-Eddington-ratio nuclei, strongly host-dominated systems, and sources with very weak broad lines. Third, because we rely on SDSS spectroscopy for uniformity, the selection is limited by the SDSS aperture, seeing, spectral resolution, and signal-to-noise ratio. In low-mass galaxies, where the AGN continuum and broad lines are often strongly diluted by host-galaxy light, uncertainties in continuum subtraction and stellar absorption modeling can become comparable to the broad-line signal itself. Therefore, the present sample should be regarded as a high-purity, but incomplete, sample of SDSS-era broad-line IMBH candidates, rather than a representative census of the underlying IMBH population.

RGG~118, a well-known IMBH candidate \citep{Reines13,Baldassare15}, provides a clear illustration of our selection limitations. Its broad H$\alpha$ component is extremely weak and lies close to the practical detection boundary of SDSS-like spectra. In our analysis, RGG~118 fails the automatic selection because its broad H$\alpha$ component is too weak relative to the narrow H$\alpha$ line, and its integrated broad-line signal is not sufficiently significant. Therefore, it would not be securely recovered by the automatic cuts alone, and was retained only after careful visual inspection, aided by its previous confirmation as a robust IMBH candidate. This example illustrates that other RGG~118-like systems may have been missed, especially when the nuclear emission is more strongly diluted by host-galaxy light or when the spectra have lower SNR. Moving beyond this SDSS selection boundary will require higher-quality spectroscopy and improved nuclear--host separation. DESI provides a moderate improvement over SDSS for weak broad-line measurements (see \S\ref{sec:desi}), while our ongoing search for AGNs in dwarf galaxies in the North Ecliptic Pole (NEP) field with the Subaru Prime Focus Spectrograph (PFS) will further push toward lower-contrast and lower-luminosity nuclear activity (Ren et al., in prep.).

\subsection{BH mass uncertainty}\label{sec:BHerr}

The black hole masses in this work are based on single-epoch broad H$\alpha$ measurements and therefore carry both statistical and systematic uncertainties. The formal uncertainties arise from the measurements of the broad-H$\alpha$ luminosity and line width, with the latter being particularly important because the virial mass depends nearly quadratically on FWHM. This problem is amplified for weak broad-line sources, where imperfect stellar-continuum subtraction, residual Balmer absorption, narrow-line wings, and blending with the adjacent [\ion{N}{2}] lines can affect the recovered broad H$\alpha$ flux and FWHM. These effects are most severe for host-dominated spectra and for objects close to the broad-line detection limit.

A further, and potentially dominant, systematic uncertainty arises from the adopted broad-line profile and its physical interpretation. We model broad H$\alpha$ with one or two Gaussian components, whereas real BLR profiles may be asymmetric, disk-like, Lorentzian, or contain non-virial wings. For example, \citet{VeronCetty01} showed that in NLS1s the broad Balmer lines are often better described by Lorentzian rather than Gaussian profiles; fitting them with Gaussians can overestimate the FWHM by a factor of $\sim1.5$--$2.5$, corresponding to a systematic black hole mass overestimate of $\sim0.36$--$0.80$ dex for virial estimators. A related but physically distinct concern is that part of the line width may not trace virial motion. In NGC~4395, \citet{Laor06} detected highly symmetric exponential H$\alpha$ wings extending to $\sim2500\ {\rm km\ s^{-1}}$ and attributed them to electron scattering within the photoionized BLR gas. Although the scattering wings have little effect on the FWHM in NGC~4395, they could produce larger biases in denser systems \citep[e.g., Little Red Dots, LRDs, at high redshift,][]{Rusakov26}.

These limitations highlight the necessity of direct measurements using RM. Single-epoch virial masses typically have an intrinsic uncertainty of $\sim0.4$--$0.5$ dex relative to RM-calibrated masses, owing to the use of an empirical $R$--$L$ relation, a single-epoch luminosity, and a static line profile \citep[e.g.,][]{Shen13,Peterson14}. In contrast, well-sampled RM campaigns directly measure the BLR radius from the time lag and use the variable broad-line component to define the virial velocity, reducing the uncertainty of the virial product to typically $\sim0.1$--$0.2$ dex \citep{Peterson04}, although the absolute mass scale still depends on the virial factor. While dynamical modeling of velocity-resolved RM data enables direct black hole mass measurements for individual AGNs without relying on an externally calibrated virial factor \citep{Pancoast14}. A small but precise set of RM-based IMBH masses will therefore provide the necessary anchor for extending the $R$--$L$ relation, the $M_{\rm BH}$--$M_\ast$ relation, and the $M_{\rm BH}$--$\sigma_\ast$ relation toward lower black hole masses, and will lay the foundation for more reliable mass estimates in much larger IMBH samples.

\begin{figure}[ht]
    \centering
  \includegraphics[width=1\linewidth]{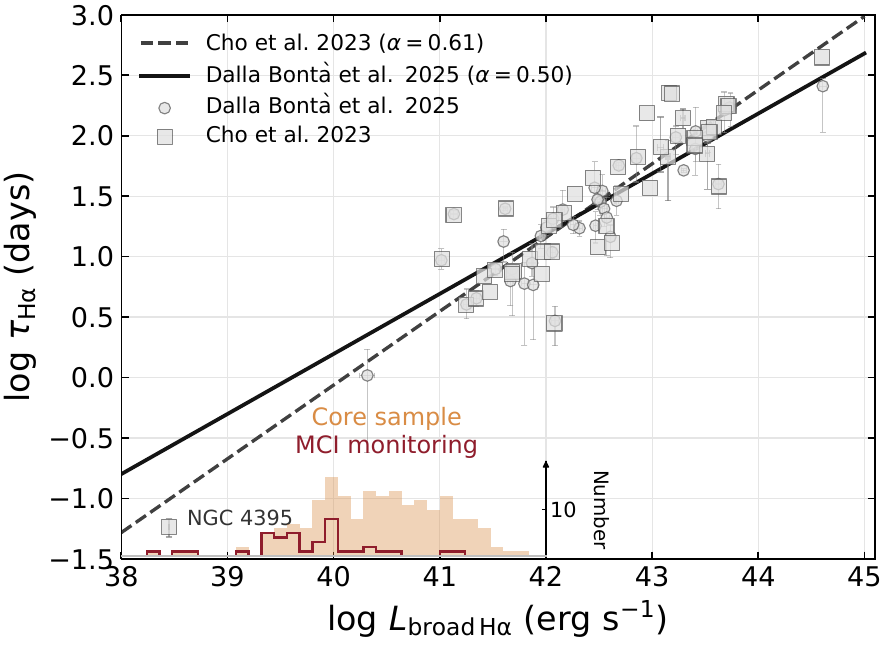}
    \caption{
Broad H$\alpha$ \(R\)–\(L\) relation. Grey circles and squares represent the H$\alpha$ RM samples from \citet{Dallabonta25} and \citet{Cho23}, respectively; NGC~4395, included in the Cho23 compilation, marks the current low-luminosity anchor. The inset shows the broad H$\alpha$ luminosity distribution of our 192 core sample and the MCI subset, which extend the coverage toward lower luminosities.
}
    \label{fig:R-L}
\end{figure}

\subsection{Extending the BLR $R-L$ Relation into the IMBH Regime}
The BLR $R-L$ relation provides the foundation for estimating black hole masses from single-epoch spectra, by using reverberation-mapped AGNs to calibrate virial mass estimators for much larger samples. While the optical H$\beta$ $R-L$ relation is well established \citep[e.g.,][]{Bentz13}, H$\beta$ is often difficult to use for IMBHs because the broad component is weak. Broad H$\alpha$ therefore offers a more practical route for low-mass, low-luminosity AGNs, building on H$\alpha$-based single-epoch mass calibrations \citep[e.g.,][]{Greene05} and recent direct H$\alpha$ reverberation measurements.

Recent H$\alpha$ $R-L$ calibrations, however, still show appreciable differences at the low-luminosity end. \citet{Cho23} combined five reliable new H$\alpha$ lags from the SAMP reverberation program with previously 42 published H$\alpha$ measurements, including the low-luminosity anchor NGC 4395, and reported a relatively steep slope of $\alpha=0.61 \pm 0.04$. In contrast, \citet{Dallabonta25} constructed a larger and more homogeneous H$\alpha$ RM sample by remeasuring H$\alpha$ lags and broad-line luminosities for 42 objects drawn from the literature and recent RM programs, finding a shallower slope of $\alpha=0.50 \pm 0.02$. As shown in Figure~\ref{fig:R-L}, the two relations are broadly consistent over the luminosity range occupied by most existing RM AGNs, but diverge toward low luminosities. Below $\log (L_{\rm H\alpha}/{\rm erg\,s^{-1}}) \lesssim 40$, the predicted H$\alpha$ lags differ by $\gtrsim0.25$ dex. This low-luminosity regime is currently anchored largely by NGC 4395, leaving substantial uncertainty in whether the H$\alpha$ $R-L$ relation remains close to the photoionization expectation or steepens toward the IMBH regime. Our sample extends the luminosity coverage into this poorly constrained low-$L_{\rm H\alpha}$ regime and can provide direct leverage on the H$\alpha$ $R-L$ relation for low-mass black holes.

\subsection{Comparison with DESI} \label{sec:desi} 
DESI and SDSS offer complementary advantages for detecting broad H$\alpha$ in active IMBHs. DESI provides a larger aperture, smaller fibers, and higher spectral resolution, improving nuclear contrast and line-profile separation, whereas the longer SDSS exposures and larger fibers collect more total light and often yield adequate spectral SNR, at the cost of stronger host-galaxy dilution. Applying host-subtracted multi-component emission-line fitting to DESI DR1 spectra, \citet{Pucha26} reported a larger sample of broad AGN candidates extending to very low virial black-hole masses. They claimed 792 objects with $\log(M_{\rm BH}/M_\odot)<6$ and 50 objects with $\log(M_{\rm BH}/M_\odot)<5$, reaching down to $\sim10^{4.4}\,M_\odot$. These low-mass objects all come from the AGN-dominated or composite regions of the BPT diagram. Cross-matching the DESI catalog of \citet{Pucha26} with our catalog yields 38 overlapping IMBH candidates with $M_{\rm BH}<10^6\,M_\odot$ and available SDSS spectra. Of these, 28/38 (73.7\%) satisfy our broad-line criteria, while the remaining 10 are excluded: 4 due to low spectral SNR, 5 because their broad-line widths are comparable to those of the narrow lines, and 1 due to outflow contamination. The DESI sample similarly shows no apparent turnover at the low-mass end, consistent with our results in Figure~\ref{fig:relations}, broadly favoring a light-seed scenario with a high black hole occupation fraction in present-day low-mass galaxies.

\begin{figure*}[ht]
    \centering
  \includegraphics[width=1\linewidth]{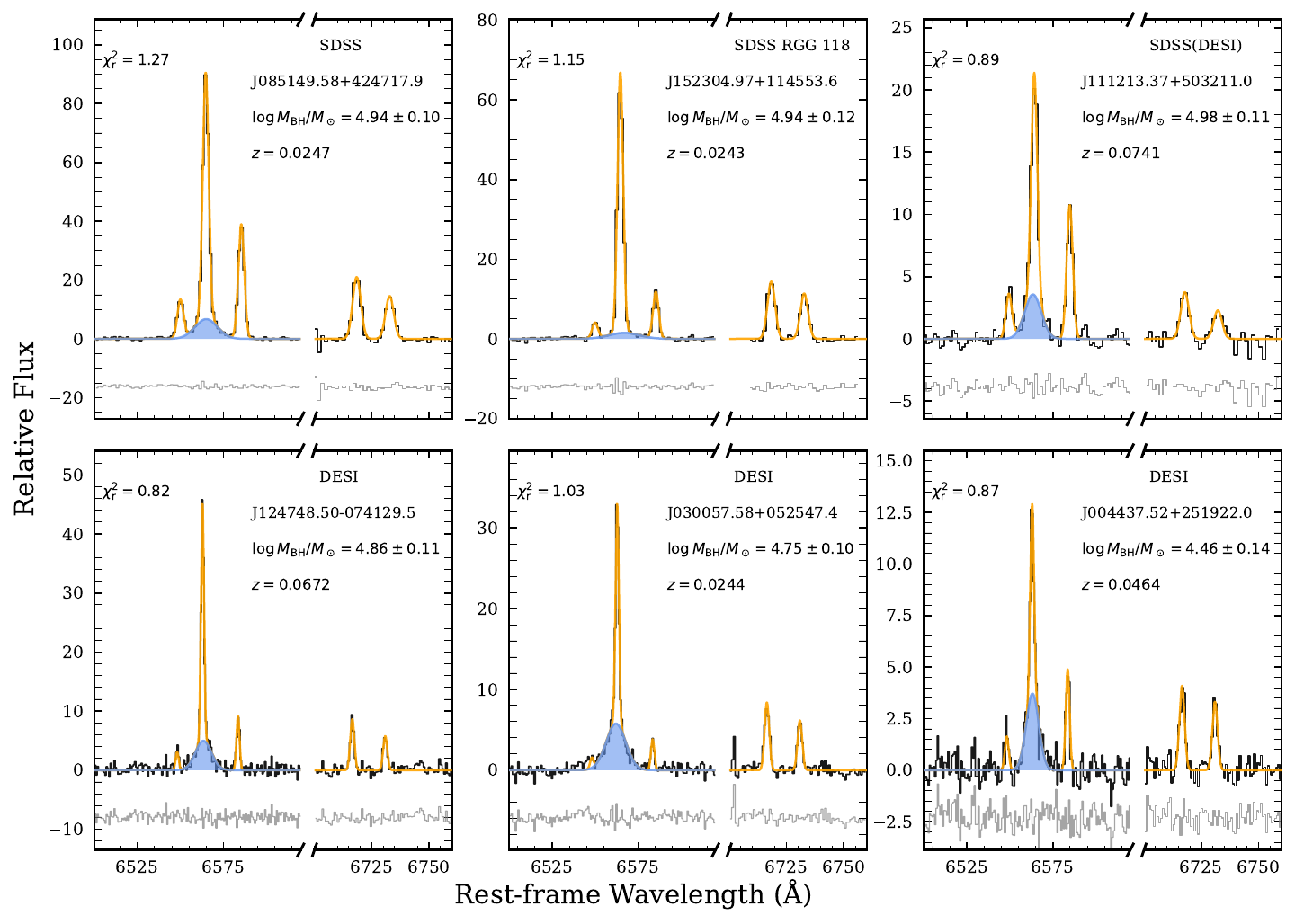}
    \caption{Spectral decomposition of the H$\alpha$ regions for six IMBH candidates with $\log(M_{\rm BH}/M_\odot)<5$. {\bf Top panels:} three sources selected from SDSS spectra in this work, including one also identified by DESI; NGC~4395, another source in this mass range, is shown in Figure~\ref{fig:fit_examples}. RGG~118 is highlighted as an example with an extremely weak broad H$\alpha$ component \citep{Baldassare15}. {\bf Bottom panels:} three additional robust DESI candidates from \citet{Pucha26}; together with the overlapping source in the upper-right panel, they comprise the four most robust DESI candidates. All sources with available X-ray coverage are detected; the final two DESI objects lack X-ray observations. Black, orange, blue, and gray denote the observed spectra, best-fitting models, broad H$\alpha$ components, and residuals, respectively.}
    \label{fig:6BHs}
\end{figure*}

Because the lowest-mass candidates are particularly difficult to validate, we reanalyzed all 50 DESI sources with $\log(M_{\rm BH}/M_\odot)<5$ reported by \citet{Pucha26} using our spectral fitting, selection criteria, and visual inspection. We identified four robust DESI candidates with clean broad H$\alpha$ components and no obvious outflow contamination. Figure~\ref{fig:6BHs} compares these four objects with three previously reported SDSS-selected low-mass sources, including one source common to both samples. All three SDSS-selected sources have X-ray counterparts. Among the four DESI candidates, two are X-ray detected, including the newly identified source J1247$-$0741 with $\log L_{\rm X,0.2-2.3\,keV}=41.90^{+0.13}_{-0.19}\ {\rm erg\ s^{-1}}$ in eROSITA, while the remaining two lack current X-ray coverage. Compared with the SDSS-selected sources, the DESI candidates have narrower broad H$\alpha$ lines, with mean widths of 510 ~km~s$^{-1}$ and 730~km~s$^{-1}$, respectively, and lower-mass hosts, with mean $\log(M_\star/M_\odot)=9.02$ versus 9.58. The full DESI IMBH sample likewise has host masses lower by $\sim0.2$ dex than our SDSS sample, indicating that DESI preferentially reaches fainter and less massive galaxies. However, despite its higher spectral resolution, DESI does not extend substantially below the SDSS black hole mass limit, highlighting that the reliable identification of genuine broad-line sources remains the main challenge at the lowest masses. Subaru/PFS, with an 8.2-m aperture and substantially greater collecting area, will test whether improved sensitivity can overcome this limitation and reveal a much larger population of lower-mass active black holes.

\subsection{BH Seed-formation Mechanisms}
JWST has uncovered a substantial population of broad-line AGNs, including LRDs, at $z\gtrsim4$ whose inferred black-hole masses often lie $1$--$3$ dex above the local $M_{\rm BH}$--$M_\star$ relation \citep{Harikane23,Uebler23,Maiolino24,Furtak24}, although electron-scattering broadening may lead to some overestimation \citep{Rusakov26}. At face value, these overmassive black holes appear to favor massive seeds formed through direct collapse, as such seeds can more readily reach the observed masses within the limited cosmic time available. However, current high-redshift samples are strongly flux- and broad-line selected, preferentially detecting the most massive and rapidly accreting black holes at fixed host mass, while fainter, lower-mass, and lower-Eddington-ratio systems remain largely inaccessible. Measurement scatter and uncertainties in both black-hole and host-galaxy masses may further enhance the apparent offsets from the local relation \citep{Lauer07,Li25}.

In contrast, we find no compelling overmassive black holes in our local active sample, with $M_{\rm BH}/M_\star\lesssim0.007$ (Figure~\ref{fig:relations}). Both the $M_{\rm BH}$--$M_\star$ and $M_{\rm BH}$--$\sigma_\star$ distributions extend smoothly down to $\log(M_{\rm BH}/M_\odot)\simeq5$, without evidence for a low-mass turnover, consistent with previous local broad-line AGN studies \citep{Reines15,Pucha25,Pucha26}. Our active IMBH candidates instead appear mildly undermassive relative to their hosts, albeit with substantial mass uncertainties. An independent dynamical example outside our sample is NGC~205, for which \citet{Nguyen19} reported $M_{\rm BH}=6.8^{+31.9}_{-2.2}\times10^3\,M_\odot$ in a quiescent galaxy with $M_\star=9.7\times10^8\,M_\odot$, corresponding to $M_{\rm BH}/M_\star\simeq7\times10^{-6}$. Its best-fitting mass lies below the canonical direct-collapse seed range, although the uncertainty overlaps its lower end. The existence of such low-mass and undermassive black holes, together with the smooth continuation of both scaling relations below $10^5\,M_\odot$, disfavors seeding models with an exclusive mass floor near the canonical heavy-seed scale and requires a substantial, and possibly dominant, contribution from lighter seeds, particularly if the black hole occupation fraction remains high in low-mass galaxies. This inference assumes, however, that subsequent accretion, mergers, and selection effects have not completely erased or distorted the initial seed signatures. The apparent contrast with the overmassive black holes reported by JWST at high redshift may partly reflect subsequent black hole--galaxy evolution, because some present-day IMBHs or our active IMBHs may no longer be pristine seed relics.

\section{Conclusions and Prospects}\label{sec:conclusion}

In this work, we compile 1447 previously reported low-mass broad-line AGN candidates, primarily from SDSS-era spectroscopic studies (Table \ref{tab:sample}). For the 1360 sources with available SDSS spectra, we perform a uniform spectral decomposition, reassess the significance of their broad H$\alpha$ components, and recompute their single-epoch black hole masses using a consistent virial calibration \citep{Reines13}. This homogeneous analysis provides a crucial sample for characterizing the local active IMBH population and investigating black hole--galaxy scaling relations. Our main conclusions are summarized as follows.

\begin{enumerate}

\item We identify 192 robust IMBH candidates at $z\lesssim0.3$ with $\log(M_{\rm BH}/M_\odot)<6$, including four particularly compelling sources with $\log(M_{\rm BH}/M_\odot)<5$ (Figures~\ref{fig:fit_examples} \& \ref{fig:6BHs}; see also recent DESI results). This sample has a median stellar mass of $\log(M_\star/M_\odot)=10.16 \pm 0.47$ and only 19 sources are hosted by dwarf galaxies with $\log(M_\star/M_\odot)<9.5$. The median black hole-to-stellar mass ratio is $M_{\rm BH}/M_\star=3.7\times10^{-5}$, suggesting that these systems may host black holes that are undermassive relative to the stellar masses of their host galaxies, although substantial uncertainties remain in both the black hole and stellar mass estimates.

\item Both the $M_{\rm BH}$--$M_\star$ and
$M_{\rm BH}$--$\sigma_\star$ relations extend smoothly down to
$\log(M_{\rm BH}/M_\odot)\simeq5$, with no clear evidence for a low-mass
flattening. Although recent JWST discoveries of
apparently overmassive black holes at high redshift have been interpreted as
possible evidence for heavy seeds
\citep[e.g.,][]{Kokorev23,Larson23,Maiolino24}, we find no comparably strong
heavy-seed imprint in the local scaling relations. Within current
uncertainties, this continuity may indicate a substantial contribution from
light seeds to the local low-mass black hole population. Alternatively,
subsequent black hole--galaxy evolution may have erased the initial seed
signature, such that at least some local active IMBHs are no longer pristine
seed relics.

\item Current empirical H$\alpha$ $R-L$ relations diverge substantially when extrapolated to the low-luminosity, low-mass regime: below $\log (L_{\rm H\alpha}/{\rm erg\,s^{-1}})\lesssim40$, the \citet{Cho23} and \citet{Dallabonta25} calibrations differ by $\gtrsim0.25$ dex in predicted lag respectively. New measurements are therefore needed to anchor the H$\alpha$ $R-L$ relation directly in this poorly constrained regime.

\end{enumerate}

The broad H$\alpha$ detections and single-epoch virial masses adopted here
nevertheless require further confirmation. Repeated spectroscopy will be
essential for establishing the persistence of the broad-line components over
timescales exceeding a decade (Wu et al. in prep.), while optical and
infrared variability \citep[e.g.,][]{Pan26,Sun25}, X-ray and radio detections \citep{Dong12a,YangJ23}, and broadband SEDs will provide independent evidence for accretion activity. Ultimately, RM with the future CSST/MCI will provide direct black hole mass measurements and establish robust low-mass anchors for black hole demographics and black hole--galaxy scaling relations.


\begin{table*}[htbp]
\centering
\caption{\centering The 38 robust IMBH candidates with $\log(M_{\rm BH}/M_\odot)<6$ and $r<18$ mag whose modeled broad H$\alpha$ profiles have at least 50\% of their integrated flux within the 50\%-transmission interval of the CSST/MCI F658N or F673N filters. The sources are listed in order of increasing redshift. $f_{\rm broad}^{\rm line}$ and $f_{\rm broad}^{\rm filter}$ denote the fractional contributions of broad H$\alpha$ to the total transmitted emission-line flux and total narrow band flux, respectively. The $r$-band magnitudes are from SDSS DR18 cModel photometry; sources marked with $\dagger$ use Pan-STARRS DR2 stack Kron magnitudes.  }
\label{tab:mci}
\begin{tabular}{ccccccccc}
\hline
Name & RA & Dec. & $z$ & $r$ (mag) & $\log M_{\rm BH}$ & Filter & $f_{\rm broad}^{\rm line}$ & $f_{\rm broad}^{\rm filter}$ \\
\hline
J122548.87+333248.7 & 186.4536 & 33.5469 & 0.0010 & 13.91 & 4.52 & F658N & 0.387 & 0.344 \\
J103234.85+650227.9 & 158.1452 & 65.0411 & 0.0056 & 12.74 & 5.52 & F658N & 0.427 & 0.058 \\
\hline
J120208.36+174124.9 & 180.5348 & 17.6902 & 0.0219 & 14.69 & 5.89 & F673N & 0.366 & 0.043 \\
J112545.35+240824.0 & 171.4390 & 24.1400 & 0.0237 & 13.91 & 5.80 & F673N & 0.546 & 0.171 \\
J152304.97+114553.5 & 230.7707 & 11.7649 & 0.0243 & 16.76 & 4.94 & F673N & 0.080 & 0.033 \\
J085149.59+424717.8 & 132.9566 & 42.7883 & 0.0247 & 15.53 & 4.94 & F673N & 0.161 & 0.049 \\
J112339.01+232259.1 & 170.9126 & 23.3831 & 0.0257 & 16.99 & 5.38 & F673N & 0.215 & 0.054 \\
J102911.51+390653.5 & 157.2979 & 39.1149 & 0.0260 & 14.86 & 5.94 & F673N & 0.438 & 0.060 \\
J093408.60+175644.0 & 143.5358 & 17.9456 & 0.0271 & 15.51 & 5.73 & F673N & 0.466 & 0.130 \\
J130141.56+100100.1 & 195.4232 & 10.0167 & 0.0273 & 15.34 & 5.34 & F673N & 0.220 & 0.037 \\
J004042.10$-$110957.6 & 10.1754 & $-$11.1660 & 0.0274 & 16.79 & 5.86 & F673N & 0.553 & 0.063 \\
J130456.96+395529.7 & 196.2373 & 39.9249 & 0.0274 & 14.39 & 5.95 & F673N & 0.853 & 0.214 \\
J130340.81+534323.7 & 195.9201 & 53.7232 & 0.0276 & 14.85 & 5.92 & F673N & 0.410 & 0.074 \\
J143450.62+033842.5 & 218.7109 & 3.6451 & 0.0283 & 14.91 & 5.86 & F673N & 0.479 & 0.131 \\
J121923.73+303002.7 & 184.8489 & 30.5007 & 0.0284 & 15.76 & 5.23 & F673N & 0.552 & 0.078 \\
J162539.87+404804.3 & 246.4161 & 40.8012 & 0.0286 & 17.00 & 5.29 & F673N & 0.354 & 0.072 \\
J171725.53+291107.9 & 259.3564 & 29.1855 & 0.0286 & 15.82 & 5.36 & F673N & 0.302 & 0.026 \\
J141738.88+072412.3 & 214.4120 & 7.4034 & 0.0288 & 15.89 & 5.12 & F673N & 0.183 & 0.101 \\
J142307.52+283542.3 & 215.7813 & 28.5951 & 0.0293 & 14.67 & 5.77 & F673N & 0.291 & 0.096 \\
J024912.86$-$081525.6 & 42.3036 & $-$8.2571 & 0.0296 & 15.57 & 5.70 & F673N & 0.664 & 0.169 \\
J144536.84+270205.5 & 221.4035 & 27.0349 & 0.0298 & 14.73$^\dagger$ & 5.85 & F673N & 0.457 & 0.254 \\
J172005.27+535703.0 & 260.0220 & 53.9508 & 0.0298 & 16.33 & 5.36 & F673N & 0.755 & 0.053 \\
J023310.79$-$074813.3 & 38.2950 & $-$7.8037 & 0.0310 & 15.70 & 5.71 & F673N & 0.255 & 0.032 \\
J091424.76+115625.6 & 138.6031 & 11.9404 & 0.0312 & 16.23 & 5.83 & F673N & 0.395 & 0.124 \\
J160531.85+174826.2 & 241.3827 & 17.8073 & 0.0314 & 17.84 & 5.20 & F673N & 0.485 & 0.092 \\
J080142.58+420019.4 & 120.4275 & 42.0054 & 0.0320 & 15.95 & 5.92 & F673N & 0.167 & 0.036 \\
J095418.15+471725.0 & 148.5758 & 47.2903 & 0.0330 & 16.71 & 5.14 & F673N & 0.155 & 0.084 \\
J093401.24+245342.5 & 143.5052 & 24.8951 & 0.0334 & 15.49$^\dagger$ & 5.41 & F673N & 0.382 & 0.050 \\
J161809.38+361957.8 & 244.5391 & 36.3327 & 0.0339 & 15.78 & 5.99 & F673N & 0.850 & 0.390 \\
J095540.47+050236.5 & 148.9187 & 5.0435 & 0.0340 & 14.39 & 5.33 & F673N & 0.375 & 0.073 \\
J162636.40+350242.1 & 246.6517 & 35.0450 & 0.0342 & 15.07 & 5.50 & F673N & 0.666 & 0.200 \\
J112103.70+342709.8 & 170.2654 & 34.4527 & 0.0345 & 16.57 & 5.86 & F673N & 0.492 & 0.051 \\
J120325.67+330846.2 & 180.8571 & 33.1462 & 0.0350 & 17.65 & 5.28 & F673N & 0.145 & 0.048 \\
J162938.38+384139.3 & 247.4099 & 38.6942 & 0.0356 & 15.74 & 5.80 & F673N & 0.264 & 0.150 \\
J233837.09$-$002810.5 & 354.6545 & $-$0.4696 & 0.0356 & 16.09 & 5.96 & F673N & 0.262 & 0.090 \\
J162612.35+241330.5 & 246.5514 & 24.2251 & 0.0373 & 15.93 & 5.21 & F673N & 0.335 & 0.047 \\
J144850.09+160803.2 & 222.2087 & 16.1342 & 0.0383 & 17.12 & 5.20 & F673N & 0.753 & 0.266 \\
J152637.36+065941.6 & 231.6557 & 6.9949 & 0.0384 & 17.48 & 5.51 & F673N & 0.504 & 0.185 \\
\hline
\end{tabular}
\end{table*}

\bigskip
\begin{acknowledgements}
\noindent We acknowledge support from the National Key R\&D Program of China (No.~2023YFA1607903). HXG acknowledges support from the NSFC (Nos.~12473018 and 12522304) and the Overseas Center Platform Projects, CAS (No.~178GJHZ2023184MI); MYS from the NSFC (No.~12322303) and the Natural Science Foundation of Fujian Province of China (No.~2022J06002); LCH from the NSFC (No.~12233001) and the China Manned Space Program (CMS-CSST-2025-A09); and ACG partially from the CAS President's International Fellowship Initiative (PIFI; No.~2026PVA0040). MFG is supported by the Shanghai Pilot Program for Basic Research--Chinese Academy of Science, Shanghai Branch (No.~JCYJ-SHFY-2021-013), the National SKA Program of China (No.~2022SKA0120102), the science research grants from the China Manned Space Project (No.~CMSCSST-2021-A06), and the Original Innovation Program of the Chinese Academy of Sciences 715 (No.~E085021002). RSL acknowledges support from the NSFC (No.~12325302), and JXW from GSFC (No.~12533006) and the Guizhou Provincial Major Scientific and Technological Program XKBF (2025)010 and XKBF (2025)011. ABK, DI, and L\v{C}P acknowledge funding from the University of Belgrade--Faculty of Mathematics (Contract No.~451-03-33/2026-03/200104) and the Astronomical Observatory Belgrade (Contract No.~451-03-33/2026-03/200002), through grants from the Ministry of Science, Technological Development and Innovation of the Republic of Serbia. XHD acknowledges support from the National Natural Science Foundation of China (No.~12573017). MM acknowledges support from the Spanish Ministry of Science and Innovation through project PID2024-159201NB-C22. This work was also partly supported by the Spanish program Unidad de Excelencia Mar\'ia de Maeztu CEX2020-001058-M, financed by MCIN/AEI/10.13039/501100011033, and the MaX-CSIC Excellence Award MaX4-SOMMA-ICE. WWZ and XBW acknowledge the support from the NSFC grant (No.~12633003)

\end{acknowledgements}

\facility{SDSS, DESI}
\software{AstroPy \citep{Astropy2018}, PyQSOFit \citep{Guo18}}

\bibliography{ms}{}
\bibliographystyle{aasjournalv7}

\appendix
\setcounter{figure}{0}
\renewcommand{\thefigure}{A\arabic{figure}}

\section{Comparison of Derived AGN Parameters} \label{app:comparison}
As a consistency check, we compare our measurements of the broad H$\alpha$ FWHM and luminosity, as well as the resulting single-epoch black hole masses, with values reported in the literature. For the 675 confident broad-line candidates with $\log(M_{\rm BH}/M_\odot)<6.5$ (192 core and 483 comparison sources), the median offsets are small, indicating no strong global systematic bias. However, the substantial scatter around these median offsets shows that measurements for individual objects can differ considerably among studies, likely because of heterogeneous spectral decomposition, line-profile modeling, and mass calibrations. This source-to-source dispersion highlights the value of uniformly reanalyzing the full sample with a consistent fitting procedure, while recognizing that our adopted methodology is not necessarily definitive.

\section{AGN luminosity and stellar velocity dispersion}\label{app:AppB}
Thanks to the extensive, high-quality spectroscopic and imaging data assembled by Greene and Ho and their collaborators, we are able to place our measurements on a well-established observational footing and directly assess their reliability.

\begin{figure*}
    \centering
    \includegraphics[width=1\linewidth]{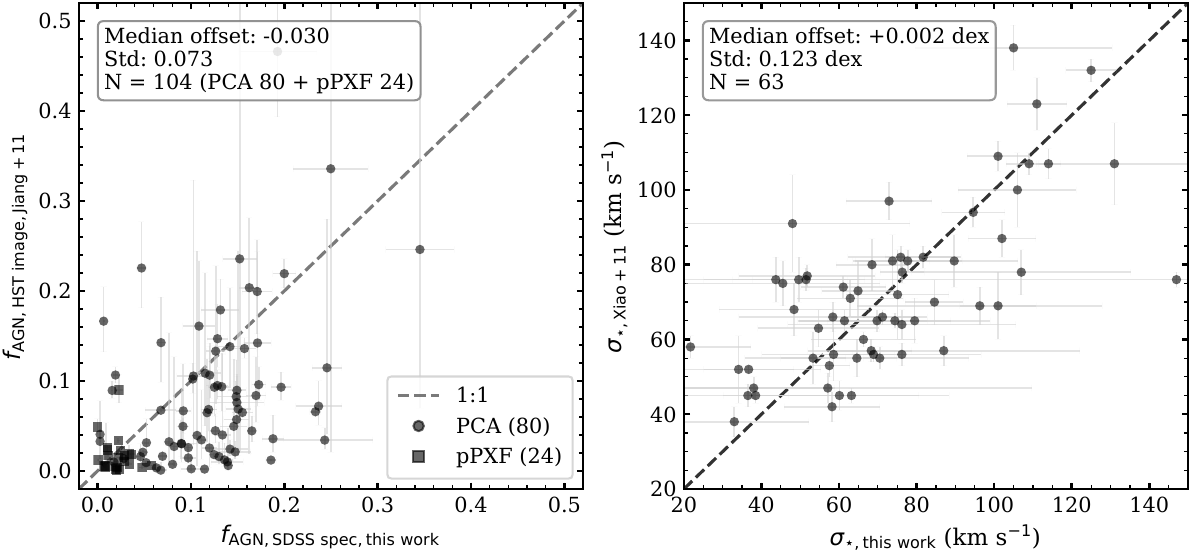}
    \caption{Comparison of AGN fraction and stellar velocity dispersion. {\bf Left panel:} Comparison of AGN fractions from SDSS spectral decomposition and HST F814W imaging decomposition \citep{Jiang11}. Circles and squares indicate PCA- and pPXF-based host decompositions, respectively. {\bf Right panel:} Stellar velocity dispersions measured from our SDSS spectral decomposition compared with high-resolution measurements from \citet{Xiao11} using Keck/ESI and Magellan/MagE spectra. Median offsets are defined as Y relative to X; scatters and sample sizes are indicated in each panel. }
    \label{fig:L&sigma}
\end{figure*}

We decompose the SDSS spectra of 145 IMBH candidates cross-matched with the sample of \citet{Jiang11} into host-galaxy and AGN components using the method described in Section~\ref{sec:decomp}. Successful decompositions at rest-frame 8140\,\AA, measured over 8115--8165\,\AA, are obtained for 104 sources. To assess the reliability of the spectroscopic decomposition, we compare the resulting AGN fractions with those independently derived by \citet{Jiang11} from {\tt GALFIT} modeling of the HST/WFPC2 F814W images. In their analysis, the AGN is represented by an unresolved point source, while the host galaxy is modeled as a S{\'e}rsic bulge plus an exponential disk. Uncertainties in the imaging-based AGN fractions are estimated through Monte Carlo sampling of the reported magnitude errors. As shown in Figure~\ref{fig:L&sigma}, the spectroscopic AGN fractions are systematically higher than the imaging-based values, with median fractions of 0.102 and 0.035, respectively, corresponding to an offset of a factor of $\sim3$. The same trend is found in 78 of the 104 sources (75\%). This comparison suggests that the spectroscopic decomposition commonly assigns some residual host-galaxy or circumnuclear emission to the AGN continuum, causing its luminosity to be overestimated. A similar bias was reported by \citet{LiYX21} for continuum luminosities derived from ground-based spectra. Such spectroscopic AGN continuum luminosities should therefore generally be regarded as upper limits to the intrinsic nuclear luminosities. The inferred black hole masses should be considerably less affected, however, because they are based on the luminosity and width of the broad H$\alpha$ component, for which contamination from host-galaxy emission is relatively minor.

On the other hand, the stellar velocity dispersions ($\sigma_*$) derived in this work show good overall agreement with the high-resolution measurements of \citet{Xiao11}. Our $\sigma_*$ values and their uncertainties are obtained by fitting the host-galaxy component with pPXF. The comparison is based on 63 objects with valid $\sigma_*$ measurements and uncertainties in both our catalog and the pre-matched \citet{Xiao11} catalog. As shown in the right panel of Figure~\ref{fig:L&sigma}, although the typical uncertainties of the SDSS-based measurements are approximately a factor of $\sim$3.1 larger than those reported by \citet{Xiao11}, no significant systematic offset is found, with a median offset of $\log(\sigma_{\rm Xiao+11}/\sigma_{\rm this\,work})=0.002$ dex. The larger uncertainties are expected because our measurements are based on global spectral fitting of SDSS low-resolution spectra, whereas \citet{Xiao11} measured $\sigma_*$ from the Ca\,{\sc ii} triplet absorption features around 8500 \AA\ in high-resolution Keck/ESI and Magellan/MagE spectra, which provide a more direct and precise tracer of the stellar velocity dispersion. Nevertheless, the good agreement between the two datasets indicates that our velocity dispersion measurements are robust across the full sample.

\begin{figure*}[h]
    \centering
    \includegraphics[width=1\linewidth]{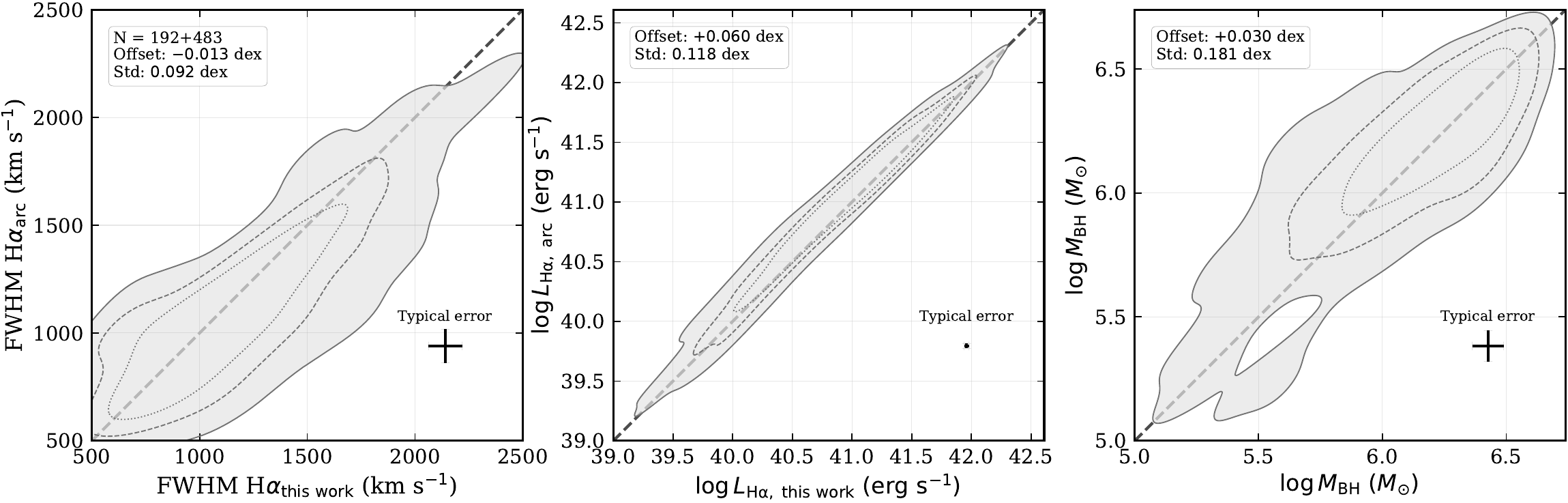}
    \caption{Comparison of our broad H$\alpha$ FWHM, luminosity, and inferred black hole masses with literature measurements for confident broad-line IMBH candidates with $\log(M_{\rm BH}/M_\odot)<6.5$. Dashed lines indicate equality, contours trace the sample density of 60\%, 80\% and 95\%, and the error bar shows the typical uncertainty. The small median offsets indicate no significant systematic differences. Median offsets are defined as Y relative to X.} 
    \label{fig:com_3p}
\end{figure*}

\begin{table*}[htbp]
\centering
\caption{X-ray statistics for 1360 sources with valid black hole masses.}
\label{tab:xray}
{
\footnotesize
\setlength{\tabcolsep}{6pt}
\begin{tabular}{lcccccccccc}
\hline\hline
Survey & Band & $R_{\rm match}$ & \multicolumn{3}{c}{$\mathrm{tag}_{\rm br}=0$} & \multicolumn{3}{c}{$\mathrm{tag}_{\rm br}=1$} & Total & Det. Rate \\
\cline{4-6}\cline{7-9}
 & (keV) & ($^{\prime\prime}$) & $<6$ & $6$--$6.5$ & $\geq6.5$ & $<6$ & $6$--$6.5$ & $\geq6.5$ & & (\%) \\
\hline
XMM-Newton & 0.2--12.0 & 10 & 17 (4.8\%) & 6 (14.6\%) & 6 (11.5\%) & 32 (16.7\%) & 73 (15.1\%) & 22 (9.2\%) & 156 & 11.5\% \\
Chandra & 0.5--7.0 & 2 & 16 (4.5\%) & 3 (7.3\%) & 2 (3.8\%) & 53 (27.6\%) & 62 (12.8\%) & 16 (6.7\%) & 152 & 11.2\% \\
eROSITA & 0.2--2.3 & 10 & 12 (3.4\%) & 5 (12.2\%) & 7 (13.5\%) & 53 (27.6\%) & 122 (25.3\%) & 64 (26.7\%) & 263 & 19.3\% \\
\hline
Unique objects & -- & -- & 35 (9.9\%) & 12 (29.3\%) & 14 (26.9\%) & 100 (52.1\%) & 207 (42.9\%) & 86 (35.8\%) & 454 & 33.4\% \\
\hline\hline
\end{tabular}
\begin{minipage}{\textwidth}
\textit{Note.} Percentages are relative to the total number of sources in each $\mathrm{tag}_{\rm br}$ and $\log(M_{\rm BH}/M_\odot)$ bin. The corresponding totals are 352, 41, and 52 for $\mathrm{tag}_{\rm br}=0$, and 192, 483, and 240 for $\mathrm{tag}_{\rm br}=1$ in the $<6$, $6$--$6.5$, and $\geq6.5$ mass bins, respectively. Total detection and its rate are computed using 1360 sources with valid black hole masses. ``Unique objects'' denotes sources detected in at least one survey after deduplication. Detections are defined by positive X-ray luminosities in the merged catalog.
\end{minipage}
}
\end{table*}

\begin{deluxetable*}{@{}lcccccc@{}}
\label{tab:radio_surveys}
\scriptsize
\tablecaption{Summary of radio surveys and cross-matching results for the 1360 sources with valid black hole masses.\label{tab:radio_surveys}}
\tablewidth{0pt}
\tablehead{
\colhead{Survey} &
\colhead{Sky coverage} &
\colhead{Frequency} &
\colhead{Sensitivity} &
\colhead{Resolution} &
\colhead{R$_{\rm match}$} &
\colhead{N$_{\rm match}$} \\
&
&
\colhead{(GHz)} &
\colhead{(mJy beam$^{-1}$)} &
\colhead{($''$)} &
\colhead{($''$)} &
}
\startdata
LoTSS DR3 & $\delta \gtrsim 0^\circ$ & 0.144 & 0.092 & 6 & 5 & 546 \\
FIRST & $>$10,000 deg$^{2}$ & 1.4 & 0.15 & 5 & 5 & 90 \\
VLASS Ep.1 & $\delta>-40^\circ$ & 3 & 0.12 & 2.5 & 2 & 64 \\
VLASS Ep.2 & $\delta>-40^\circ$ & 3 & 0.12 & 2.5 & 2 & 71 \\
VLASS Ep.3 & $\delta>-40^\circ$ & 3 & 0.12 & 2.5 & 2 & 62 \\
RACS-low & $\delta\!\lesssim\!+30^\circ$ & 0.89 & 0.20 & 15--25 & 15 & 64 \\
RACS-mid & $\delta\!\lesssim\!+50^\circ$ & 1.37 & 0.20 & 9 & 8 & 96 \\
RACS-high & $\delta\!\lesssim\!+50^\circ$ & 1.66 & 0.20 & 8 & 8 & 72 \\
\enddata
\tablecomments{
Positional cross-matching was performed using the closest radio source within a fixed radius. $N_{\rm match}$ is calculated for the 1360 sources with valid black hole masses, matching the sample used in Figure~\ref{fig:xray_radio}. The reference of each surveys are: LoTSS DR3: \citet{Shimwell26}; FIRST: \citet{Becker95}; VLASS: \citet{Lacy20}; RACS-low: \citet{Hale21}; RACS-mid: \citet{Duchesne23}; RACS-high: \citet{Duchesne25}.
}
\end{deluxetable*}

\startlongtable
\begin{deluxetable*}{@{}rllll@{}}
\tabletypesize{\scriptsize}
\tablecaption{IMBH catalog data model.\label{tab:imbh_catalog_data_model}}
\tablewidth{0pt}
\tablehead{
\colhead{No.} &
\colhead{Column Name} &
\colhead{Type} &
\colhead{Units} &
\colhead{Description}
}
\startdata
1 & \texttt{ID} & int64 & -- & Row index in this table (1-based, after sorting by RA) \\
2 & \texttt{Name} & string & -- & Object name \\
3 & \texttt{RA} & float64 & deg & Right ascension (J2000) in degrees \\
4 & \texttt{DEC} & float64 & deg & Declination (J2000) in degrees \\
5 & \texttt{z} & float64 & -- & Redshift \\
6 & \texttt{logM\_star} & float64 & $M_\odot$ & Stellar mass (log10) \\
7 & \texttt{logM\_star\_err} & float64 & $M_\odot$ & Uncertainty on stellar mass \\
8 & \texttt{logSFR} & float64 & $M_\odot\,\mathrm{yr}^{-1}$ & Star formation rate (log10) \\
9 & \texttt{logSFR\_err} & float64 & $M_\odot\,\mathrm{yr}^{-1}$ & Uncertainty on star formation rate \\
10 & \texttt{logL\_Ha\_arc} & float64 & $\mathrm{erg\,s}^{-1}$ & \ha\ archival line luminosity (log10) \\
11 & \texttt{FWHM\_Ha\_arc} & int64 & $\mathrm{km\,s}^{-1}$ & \ha\ archival FWHM \\
12 & \texttt{logM\_BH\_arc} & float64 & $M_\odot$ & Archival black hole mass (log10) \\
13 & \texttt{Ref} & string & -- & Reference \\
14 & \texttt{Plate} & int32 & -- & SDSS plate identifier \\
15 & \texttt{MJD} & int32 & -- & SDSS modified Julian date \\
16 & \texttt{Fiber} & int32 & -- & SDSS fiber identifier \\
17 & \texttt{Morphology} & string & -- & SDSS morphology: point or extended \\
18 & \texttt{Re\_arcsec} & float64 & arcsec & SDSS r-band Petrosian half-light radius \\
19 & \texttt{Re\_arcsec\_err} & float64 & arcsec & Uncertainty in Re\_arcsec \\
20 & \texttt{Re\_kpc} & float64 & kpc & Physical half-light radius \\
21 & \texttt{Re\_kpc\_err} & float64 & kpc & Uncertainty in Re\_kpc \\
22 & \texttt{SDSS\_gmag} & float64 & mag & SDSS g-band cModel magnitude \\
23 & \texttt{SDSS\_gmag\_err} & float64 & mag & Uncertainty in SDSS gmag \\
24 & \texttt{SDSS\_rmag} & float64 & mag & SDSS r-band cModel magnitude \\
25 & \texttt{SDSS\_rmag\_err} & float64 & mag & Uncertainty in SDSS rmag \\
26 & \texttt{tag\_ppxf} & int16 & -- & Flag for pPXF-based host decomposition (if yes, 1)\\
27 & \texttt{tag\_br} & int64 & -- & Flag for confident broad component in \ha\ (if yes, 1) \\
28 & \texttt{frac\_host\_5100} & float64 & -- & Host-galaxy fraction at 5100 A from spectral decomposition \\
29 & \texttt{SNR\_host} & float64 & -- & S/N for host continuum \\
30 & \texttt{SNR\_Ha} & float64 & -- & S/N for broad \ha \\
31 & \texttt{Sigma\_star} & float64 & $\mathrm{km\,s}^{-1}$ & Stellar velocity dispersion \\
32 & \texttt{Sigma\_star\_err} & float64 & $\mathrm{km\,s}^{-1}$ & Uncertainty on stellar velocity dispersion \\
33 & \texttt{logL\_bol} & float64 & $\mathrm{erg\,s}^{-1}$ & Bolometric luminosity = 9.26 * L\_5100\_AGN (log10) \\
34 & \texttt{logL\_bol\_err} & float64 & $\mathrm{erg\,s}^{-1}$ & Uncertainty on logL\_bol from continuum S/N (log10) \\
35 & \texttt{logL\_5100} & float64 & $\mathrm{erg\,s}^{-1}$ & Total continuum luminosity at 5100 A (log10) \\
36 & \texttt{logL\_5100\_AGN} & float64 & $\mathrm{erg\,s}^{-1}$ & AGN continuum luminosity at 5100 A (log10) \\
37 & \texttt{logL\_Ha\_br} & float64 & $\mathrm{erg\,s}^{-1}$ & Broad \ha\ luminosity (log10) \\
38 & \texttt{logL\_Ha\_br\_err} & float64 & $\mathrm{erg\,s}^{-1}$ & Uncertainty on broad \ha\ luminosity \\
39 & \texttt{FWHM\_Ha\_br} & float64 & $\mathrm{km\,s}^{-1}$ & Broad \ha\ FWHM \\
40 & \texttt{FWHM\_Ha\_br\_err} & float64 & $\mathrm{km\,s}^{-1}$ & Uncertainty on broad \ha\ FWHM \\
41 & \texttt{logL\_Hb\_br} & float64 & $\mathrm{erg\,s}^{-1}$ & Broad \hb\ luminosity (log10) \\
42 & \texttt{logL\_Hb\_br\_err} & float64 & $\mathrm{erg\,s}^{-1}$ & Uncertainty on broad \hb\ luminosity \\
43 & \texttt{FWHM\_Hb\_br} & float64 & $\mathrm{km\,s}^{-1}$ & Broad \hb\ FWHM \\
44 & \texttt{FWHM\_Hb\_br\_err} & float64 & $\mathrm{km\,s}^{-1}$ & Uncertainty on broad \hb\ FWHM \\
45 & \texttt{logM\_BH} & float64 & $M_\odot$ & Virial black hole mass (log10) estimated from broad \ha\ based on Reines et al. (2015) \\
46 & \texttt{logM\_BH\_err} & float64 & $M_\odot$ & Uncertainty on virial black hole mass \\
47 & \texttt{Lamda\_Edd} & float64 & -- & Eddington ratio from AGN 5100 luminosity \\
48 & \texttt{Voff\_OIIIw} & float64 & $\mathrm{km\,s}^{-1}$ & [OIII]5007 wing velocity offset \\
49 & \texttt{Voff\_OIIIc} & float64 & $\mathrm{km\,s}^{-1}$ & [OIII]5007 core velocity offset \\
50 & \texttt{Sig\_OIIIw} & float64 & -- & [OIII]5007 wing significance \\
51 & \texttt{Sig\_OIIIc} & float64 & -- & [OIII]5007 core significance \\
52 & \texttt{F\_Ha\_narrow} & float64 & $\mathrm{erg\,s}^{-1}\,\mathrm{cm}^{-2}$ & Narrow \ha\ line flux \\
53 & \texttt{F\_Ha\_narrow\_err} & float64 & $\mathrm{erg\,s}^{-1}\,\mathrm{cm}^{-2}$ & Uncertainty on narrow \ha\ flux \\
54 & \texttt{F\_NII6549} & float64 & $\mathrm{erg\,s}^{-1}\,\mathrm{cm}^{-2}$ & [N II] 6549 A line flux \\
55 & \texttt{F\_NII6549\_err} & float64 & $\mathrm{erg\,s}^{-1}\,\mathrm{cm}^{-2}$ & Uncertainty on [N II] 6549 A flux \\
56 & \texttt{F\_NII6585} & float64 & $\mathrm{erg\,s}^{-1}\,\mathrm{cm}^{-2}$ & [N II] 6585 A line flux \\
57 & \texttt{F\_NII6585\_err} & float64 & $\mathrm{erg\,s}^{-1}\,\mathrm{cm}^{-2}$ & Uncertainty on [N II] 6585 A flux \\
58 & \texttt{F\_SII6718} & float64 & $\mathrm{erg\,s}^{-1}\,\mathrm{cm}^{-2}$ & [S II] 6718 A line flux \\
59 & \texttt{F\_SII6718\_err} & float64 & $\mathrm{erg\,s}^{-1}\,\mathrm{cm}^{-2}$ & Uncertainty on [S II] 6718 A flux \\
60 & \texttt{F\_SII6732} & float64 & $\mathrm{erg\,s}^{-1}\,\mathrm{cm}^{-2}$ & [S II] 6732 A line flux \\
61 & \texttt{F\_SII6732\_err} & float64 & $\mathrm{erg\,s}^{-1}\,\mathrm{cm}^{-2}$ & Uncertainty on [S II] 6732 A flux \\
62 & \texttt{F\_Hb\_narrow} & float64 & $\mathrm{erg\,s}^{-1}\,\mathrm{cm}^{-2}$ & Narrow \hb\ line flux \\
63 & \texttt{F\_Hb\_narrow\_err} & float64 & $\mathrm{erg\,s}^{-1}\,\mathrm{cm}^{-2}$ & Uncertainty on narrow \hb\ flux \\
64 & \texttt{F\_OIII4959} & float64 & $\mathrm{erg\,s}^{-1}\,\mathrm{cm}^{-2}$ & [O III] 4959 A total (core+wing) flux \\
65 & \texttt{F\_OIII4959\_err} & float64 & $\mathrm{erg\,s}^{-1}\,\mathrm{cm}^{-2}$ & Uncertainty on [O III] 4959 A flux \\
66 & \texttt{F\_OIII5007} & float64 & $\mathrm{erg\,s}^{-1}\,\mathrm{cm}^{-2}$ & [O III] 5007 A total (core+wing) flux \\
67 & \texttt{F\_OIII5007\_err} & float64 & $\mathrm{erg\,s}^{-1}\,\mathrm{cm}^{-2}$ & Uncertainty on [O III] 5007 A flux \\
68 & \texttt{F\_OII3728} & float64 & $\mathrm{erg\,s}^{-1}\,\mathrm{cm}^{-2}$ & [O II] 3728 A line flux \\
69 & \texttt{F\_OII3728\_err} & float64 & $\mathrm{erg\,s}^{-1}\,\mathrm{cm}^{-2}$ & Uncertainty on [O II] 3728 A flux \\
70 & \texttt{tag\_X-ray} & int16 & -- & 0: no observation or undetected, 1: detection in any X-ray observations \\
71 & \texttt{L\_Chandra} & float32 & $\mathrm{erg\,s}^{-1}$ & X-ray luminosity in 0.5-7.0 keV from CSC 2.1 and our analysis \\
72 & \texttt{L\_Chandra\_err\_lower} & float32 & $\mathrm{erg\,s}^{-1}$ & Lower uncertainty of the Chandra X-ray luminosity \\
73 & \texttt{L\_Chandra\_err\_upper} & float32 & $\mathrm{erg\,s}^{-1}$ & Upper uncertainty of the Chandra X-ray luminosity \\
74 & \texttt{L\_eROSITA} & float32 & $\mathrm{erg\,s}^{-1}$ & X-ray luminosity in 0.2-2.3 keV from eRASS-DR2 \\
75 & \texttt{L\_eROSITA\_err\_lower} & float32 & $\mathrm{erg\,s}^{-1}$ & Lower uncertainty of the eROSITA X-ray luminosity \\
76 & \texttt{L\_eROSITA\_err\_upper} & float32 & $\mathrm{erg\,s}^{-1}$ & Upper uncertainty of the eROSITA X-ray luminosity \\
77 & \texttt{L\_XMM} & float32 & $\mathrm{erg\,s}^{-1}$ & X-ray luminosity in 0.2--12.0 keV from 4XMM-DR14(s), XMMSL3 and our analysis \\
78 & \texttt{L\_XMM\_err\_lower} & float32 & $\mathrm{erg\,s}^{-1}$ & Lower uncertainty of the XMM-Newton X-ray luminosity \\
79 & \texttt{L\_XMM\_err\_upper} & float32 & $\mathrm{erg\,s}^{-1}$ & Upper uncertainty of the XMM-Newton X-ray luminosity \\
80 & \texttt{tag\_radio} & int64 & -- & 0: no observation or undetected, 1: detection in any radio surveys \\
81 & \texttt{F\_LoTSS} & float64 & mJy & Flux density in the LoTSS DR3 source catalog \\
82 & \texttt{F\_LoTSS\_err} & float64 & mJy & Uncertainty of the flux density in the LoTSS DR3 source catalog \\
83 & \texttt{L\_150MHz} & float64 & $\mathrm{W\,Hz}^{-1}$ & Rest-frame 150 MHz luminosity from LoTSS (log10) \\
84 & \texttt{L\_150MHz\_err} & float64 & $\mathrm{W\,Hz}^{-1}$ & Uncertainty of rest-frame 150 MHz luminosity from LoTSS \\
85 & \texttt{F\_FIRST} & float64 & mJy & Flux density in the FIRST survey \\
86 & \texttt{F\_RACS\_Low} & float64 & mJy & Flux density in the RACS-Low source catalog \\
87 & \texttt{F\_RACS\_Low\_err} & float64 & mJy & Uncertainty of the flux density in the RACS-Low source catalog \\
88 & \texttt{F\_RACS\_Mid} & float64 & mJy & Flux density in the RACS-Mid source catalogue \\
89 & \texttt{F\_RACS\_Mid\_err} & float64 & mJy & Uncertainty of the flux density in the RACS-Mid source catalog \\
90 & \texttt{F\_RACS\_High} & float64 & mJy & Flux density in the RACS-High source catalogue \\
91 & \texttt{F\_RACS\_High\_err} & float64 & mJy & Uncertainty of the flux density in the RACS-High source catalog \\
92 & \texttt{L\_1.4GHz} & float64 & $\mathrm{W\,Hz}^{-1}$ & Rest-frame 1.4 GHz luminosity from FIRST+RACS (log10) \\
93 & \texttt{L\_1.4GHz\_err} & float64 & $\mathrm{W\,Hz}^{-1}$ & Uncertainty of rest-frame 1.4 GHz luminosity from FIRST+RACS \\
94 & \texttt{F\_VLASS\_EP1} & float64 & mJy & Flux density in the VLASS Epoch 1 QL component catalog \\
95 & \texttt{F\_VLASS\_EP1\_err} & float64 & mJy & Uncertainty of the flux density in the VLASS Epoch 1 QL component \\
96 & \texttt{F\_VLASS\_EP2} & float64 & mJy & Flux density in the VLASS Epoch 2 QL component catalog \\
97 & \texttt{F\_VLASS\_EP2\_err} & float64 & mJy & Uncertainty of the flux density in the VLASS Epoch 2 QL component \\
98 & \texttt{F\_VLASS\_EP3} & float64 & mJy & Flux density in the VLASS Epoch 3 QL component catalogue \\
99 & \texttt{F\_VLASS\_EP3\_err} & float64 & mJy & Uncertainty of the flux density in the VLASS Epoch 3 QL component \\
100 & \texttt{L\_3GHz} & float64 & $\mathrm{W\,Hz}^{-1}$ & Rest-frame 3 GHz luminosity from VLASS (log10) \\
101 & \texttt{L\_3GHz\_err} & float64 & $\mathrm{W\,Hz}^{-1}$ & Uncertainty of rest-frame 3 GHz luminosity from VLASS \\
\enddata
\end{deluxetable*}

\end{document}